\documentclass[11pt,a4paper,oneside]{amsart}
\pdfoutput=1

\usepackage[a4paper,textwidth=16.2cm,textheight=23.3cm,centering]{geometry}

\usepackage[]{amsmath}
	\numberwithin{equation}{section}
\usepackage[]{amssymb}
\usepackage{amscd}
\usepackage{amsfonts}
\usepackage{mathrsfs}
\usepackage{amsthm}
	\newtheorem{thm}{Theorem}[section]

	\theoremstyle{remark}
	\newtheorem{rmk}[thm]{Remark}
	\newtheorem*{defin}{Definition}

\DeclareMathOperator{\R}{\mathbb{R}}
\DeclareMathOperator{\C}{\mathbb{C}}

\DeclareMathOperator{\ct}{\mathbb{T}}
\DeclareMathOperator{\P1}{\mathbb{P}^{1}}

\DeclareMathOperator{\Res}{\mathrm{Res}}

\newcommand{\ii}{\mathrm{i}}
\newcommand{\dd}{\mathrm{d}}
\newcommand{\mz}{\mathcal{Z}}
\newcommand{\mf}{\mathcal{F}}

\newcommand{\mC}{\mathcal{C}}
\newcommand{\cC}{\mathscr{C}}
\newcommand{\cM}{\mathscr{M}}
\newcommand{\cT}{\mathscr{T}}

\newcommand{\tr}{\mathrm{Tr}}

\usepackage[utf8]{inputenc}
\usepackage[english]{babel}

\usepackage[pdftex]{graphicx}

\usepackage[linktocpage=true,colorlinks=true,citecolor=blue,linkcolor=blue,urlcolor=blue]{hyperref}

\usepackage[toc,page]{appendix}
\usepackage{caption}
\usepackage{enumerate}

\usepackage{tikz}
		\usetikzlibrary{er,positioning,arrows.meta}

\usepackage[numbers,compress,square,comma]{natbib}

\begin{document}
\bibliographystyle{ourstyle}

	\title[]{\Large{Multiple phases and meromorphic deformations of unitary matrix models}}

	\author{Leonardo Santilli$^{\ast}$}
	\address{$^{\ast}$ Grupo de F\'{i}sica Matem\'{a}tica, Departamento de Matem\'{a}tica, Faculdade de Ci\^{e}ncias, Universidade de Lisboa, Campo Grande, Edif\'{i}cio C6, 1749-016 Lisboa, Portugal.}
	\email{lsantilli@fc.ul.pt}
	\author{Miguel Tierz$^{\dagger, \ddag}$}
	\address{$^{\dagger}$ Departamento de An\'alisis Matem\'atico y Matem\'atica Aplicada, Universidad Complutense de Madrid, 28040 Madrid, Spain}
	\email{tierz@mat.ucm.es}
	\address{$^{\ddag}$ Grupo de F\'{i}sica Matem\'{a}tica, Departamento de Matem\'{a}tica, Faculdade de Ci\^{e}ncias, Universidade de Lisboa, Campo Grande, Edif\'{i}cio C6, 1749-016 Lisboa, Portugal.}
	\email{tierz@fc.ul.pt}

		\begin{abstract}

				We study a unitary matrix model with Gross--Witten--Wadia weight function and determinant insertions. After some exact evaluations, we characterize the intricate phase diagram. There are five possible phases: an ungapped phase, two different one-cut gapped phases and two other two-cut gapped phases. The transition from the ungapped phase to any gapped phase is third order, but the transition between any one-cut and any two-cut phase is second order. The physics of tunneling from a metastable vacuum to a stable one and of different releases of instantons is discussed. Wilson loops, $\beta$-functions and aspects of chiral symmetry breaking are investigated as well. Furthermore, we study in detail the meromorphic deformation of a general class of unitary matrix models, in which the integration contour is not anchored to the unit circle. The ensuing phase diagram is characterized by symplectic singularities and captured by a Hasse diagram.

		\end{abstract}
	
	\maketitle

	\tableofcontents
\clearpage

	\section{Introduction}
	
		The study of the spectral and critical properties of models of random matrices has become a widely popular and interdisciplinary subject in this century, in great part due to the vast scope of fields where such models appear naturally and play a prominent role \cite{ForresterBook,BDSuidbook,AkeBaikDiFbook}. One among the possible many ways to highlight this relevance and versatility of random matrix theory is to simply point out the fact that the same model oftentimes appears in remarkably different contexts and, in addition, in a significant manner.\par
    A paradigmatic example of this phenomenon could very well be the so-called Gross--Witten--Wadia (GWW) model \cite{Gross:1980he,Wadia:2012fr,Wadia:1980cp}. Originally proposed in the study of gauge theory, it is ubiquitous and pivotal in many other areas, such as combinatorics, representation theory and spectral theory \cite{ForresterBook,BDSuidbook,AkeBaikDiFbook}.\par
		With this fact in mind, in this work we will study unitary matrix models, starting, precisely, with the specific case of a generalized form of the Gross--Witten--Wadia model. A possible interpretation of the model is as a one-plaquette model of two-dimensional lattice QCD with fermionic or bosonic quarks, which equivalently corresponds to a massive deformation of a model introduced by Minahan \cite{Minahan:1991pv,Minahan:1991sm}.\par
		
		We will show that, while quite simple, this model retains several features of a sensible quantum field theory in the continuum. In turn, its simplicity allows us to exploit standard techniques from random matrix theory to characterize the theory at large $N$ and suggests more general and deeper problems to consider. Some of them, we will already tackle here, by discussing at length the case of meromorphic deformations of unitary matrix model, as we explain below.\par
				Before that, a rich phase diagram will be obtained and analyzed in detail. Phase transitions such as the ones we obtain in our analysis are relevant to the study of deconfinement transitions in QCD models in four dimensions \cite{Sundborg:1999ue,Aharony:2003sx} and in black holes physics \cite{AlvarezGaume:2005fv,AlvarezGaume:2006jg,Azuma:2007fj}. More recently, this type of phase transitions has been argued to describe the critical behaviour of models exhibiting partial deconfinement \cite{Hanada:2016pwv,Berenstein:2018lrm,Hanada:2018zxn,Hanada:2019czd,Hanada:2019rzv}.\par
				We introduce the model in what follows, in Section \ref{sec:themodel}, which includes a discussion on interpretations of the model, notation, relationship with other systems and an introductory discussion of its mathematical properties, including exact evaluations without scaling limits.\par 
				Then, the main results of the paper are presented and organized as follows: there are three main contributions, as far as new results are concerned. In Section \ref{sec:Phases}, we fully characterize the rich phase structure of the unitary matrix model. In Section \ref{sec:WLinst}, we study Wilson loops in the same setting of Section \ref{sec:Phases} and discuss at length the physical interpretation of the phase transitions, including the role of instanton contributions.\par 
				
			Finally, in Section \ref{sec:merodef} we study, in a general framework and going beyond the specific model studied in the previous sections, the case where the integration contour is deformed in $\C^{\ast}$ away from the unit circle. In spite of the vast body of results on random matrix ensembles, holomorphic matrix models \cite{Lazaroiu:2003vh} are arguably understudied.\par The aim of Section \ref{sec:merodef} is to adapt the results on holomorphic matrix models to unitary matrix models. We do so for a very general set of unitary matrix models and, only as an illustrative example, we discuss the particular case of the holomorphic GWW matrix model. Non-traditional tools in this area, such as symplectic singularities and Hasse diagrams, are introduced here to fully understand the meromorphic models.\par
			We conclude with possible avenues for further research in Section \ref{sec:outlook}.
			
		\section{The model}
		\label{sec:themodel}
		In this section we present the model and study some of its exact features at finite $N$.\par

		\subsection{The model and its interpretations}
		\label{sec:modelgauge}
		
				Consider a one-plaquette model of two-dimensional lattice gauge theory \cite{Rossi:1996hs,Billo:1996pu} with gauge group $U(N)$ and $K$ pairs of real fields, that can be either bosonic or fermionic. Each pair is called a flavour. We encode the choice of matter fields in the binary variable 
				\begin{equation*}
					\epsilon = \begin{cases} +1 & \text{fermions} , \\ -1 & \text{bosons} .  \end{cases}
				\end{equation*}
				We must impose (anti-)periodic boundary conditions, so the discrete space-time is effectively reduced to a point with a loop attached to it, along which the gauge connection travels.
				Let $m_f >0$ be the mass of the $f^{\mathrm{th}}$ flavour, and introduce the notation 
				\begin{equation*}
					\mu_f ^{(\epsilon)} = 1 + \sqrt{\epsilon} m_f .
				\end{equation*}
				The partition function of this theory has the matrix model representation \cite{Minahan:1991pv} 
				\begin{equation*}
					\mz_{U(N)} ^{\epsilon, K} (\lambda  ) = \int_{U(N)} \dd U  ~ \prod_{f=1} ^{K} \left[  \det \left( \mu_f ^{(\epsilon)} - U  \right) \det \left( \mu_f ^{(\epsilon)} - U  \right)^{\dagger}  \right]^{\epsilon } ~ e^{ \frac{N}{\lambda} \left(   \tr   U +  \tr U^{\dagger} \right)  }   
				\end{equation*}
				where $\lambda \equiv N g_{\scriptscriptstyle \mathrm{YM}} ^2 $ is the 't Hooft coupling for the bare gauge coupling $g_{\scriptscriptstyle \mathrm{YM}}$, $U \in U(N)$ is the plaquette gauge variable and $\dd U$ is the normalized Haar measure on $U(N)$. Particularizing to the degenerate case with all equal masses, $\mu_f = \mu$ $\forall f=1, \dots, K$, the partition function becomes: 
				\begin{align}
					\mz_{U(N)} ^{\epsilon, K} \left( \lambda \right) &  = \int_{ U(N) }  \dd U  ~ \left[  \det \left( \mu - U  \right) \det \left( \mu  - U  \right)^{\dagger}  \right]^{ \epsilon K  }  ~ e^{ \frac{N}{\lambda} \left(   \tr   U +  \tr U^{\dagger} \right)  }  \notag \\
						& =  \oint_{\ct^{N}} \prod_{1 \le j < k \le N} \left\lvert z_j - z_k  \right\rvert^{2} ~ \prod_{j=1} ^{N} \left[  \left( \mu -z_j \right)\left( \bar{\mu} -z_j ^{-1}  \right)  \right]^{  \epsilon K } ~ e^{ \frac{N}{\lambda} \left( z_j + z_j ^{-1} \right) } ~ \frac{ \dd z_j }{ 2 \pi \ii z_j }  . \label{MM0}
				\end{align}
				The second line is written in terms of the eigenvalues $z_j \in \ct$ of $U \in U(N)$.\par
				The matrix model \eqref{MM0} is a massive deformation of the model studied in \cite{Russo:2020eif,Russo:2020puy}. Besides, \eqref{MM0} is a generalization of the celebrated Gross--Witten--Wadia (GWW) model \cite{Gross:1980he,Wadia:2012fr} by determinant insertions, and reduces to it for $K=0$ or $\mu \to \infty$. For $\lambda^{-1}=0$ the model \eqref{MM0} is a particular case of a correlator of characteristic polynomials in a Circular Unitary ensemble (CUE) \cite{haake1996secular}, a fundamental object in random matrix theory with many applications \cite{fyodorov2012freezing,fyodorov2014freezingb}. Likewise, for this value of $\lambda^{-1}=0$, it generalizes a matrix model that describes non-intersecting random walks \cite{Baik}, and has also appeared in gauge theory, for instance in \cite{Hallin:1998km} and later on in somewhat disguised forms. Still for this particular value, the model describes the probability that a random Young diagram has a value less than or equal to $N$ \cite{forrester2004application}.\par Indeed, it is well-known that there is a rich and intricate connection between problems such as vicious walkers and the study of random partitions \cite{ForresterBook,BDSuidbook}. Interestingly, the formalism of certain low dimensional gauge theories, certainly $2d$ Yang-Mills theory but also $3d$ Chern-Simons theory for example, share many common features with such, at first, seemingly different areas \cite{de2004brownian,forrester2011non}. One simple reason for this to be the case, is the well-known representation of these theories in terms of heat-kernels on group manifolds \cite{de2004brownian,forrester2011non,Romo:2011qp,gorsky2020two}.\par

				\subsection*{Integrable systems interpretation}
			
				In the two different limiting cases $K=0$ (pure GWW) and $\lambda^{-1} =0$ with $ \mu=1$, the matrix model is known to be the integral representation of a $\tau$-function of a Painlev\'{e} system, in particular Painlev\'{e} III$^{\prime}$ and V, respectively \cite{forrester2002application,forrester2004application,forrester2004discrete,adler2003recursion,ForresterBook}.\par
				In fact, the more general form of the $\tau$-function of PV is very close in form to \eqref{MM0}, differing though in having \eqref{MM0} with $e^{ \frac{N}{\lambda} z}$, instead of \eqref{MM0} with the full GWW weight.\footnote{Such an expression also appears in the study of spacing distributions \cite[Eq. (8.119)]{ForresterBook}.} If one inspects the integral representation of the $\tau$-function of PVI \cite{forrester2004application,forrester2004discrete,ForresterBook}, it is seemingly unrelated to \eqref{MM0} and that may partially explain why our model here is essentially unstudied, whereas its two limiting cases appear in many works, often analyzed simultaneously or in a comparative fashion \cite{forrester2002application,adler2003recursion,forrester2004application,forrester2004discrete,ForresterBook}.\par
				\medskip
				It is also worth mentioning another possible interpretation of \eqref{MM0}, from the point of view of integrable systems \cite{Adler99}. The study of the so-called Schur flow \cite{faybusovich1999schur,mukaihira2002schur}, analogous to Toda flows but on the unit circle, precisely entails the generalization of a given weight function of the matrix model by multiplication by the GWW weight function.\par 
				This induces a flow that has many implications. For example, the recurrence coefficients of the polynomials, orthogonal with regards to the weight function of the matrix model, satisfy the non-linear Ablowitz--Ladik equation, with the parameter $\frac{1}{g_{\scriptscriptstyle \mathrm{YM}} ^2 } = \frac{N}{\lambda}$ interpreted as time.\par
				Because of this and since our analysis is in a planar limit and centered around the matrix model, the results obtained are not obviously transferable into this integrable systems and spectral theory language \cite{nenciu2006cmv}. We will further discuss about this at the end, in the outlook Section \ref{sec:outlook}.

				\subsection*{Gauge theory interpretation}
				Rather, we consider \eqref{MM0} instead as a toy model for lattice two-dimensional QCD, although we will comment on some of the many other interpretations of the model. For example, \eqref{MM0} can be regarded as an effective description of two-dimensional QCD on a small spatial circle \cite{Hallin:1998km}. In fact, compactifying the spatial direction generates a mass gap for all but the zero-modes. Taking the small circumference limit, we are left with an effective theory with integration only over the gauge and matter zero-modes.\par
				
				It was observed in \cite{Russo:2020eif} that the massless theory with fermions, that is $\epsilon=+1$ and $\mu =1$, shows a Fisher--Hartwig (FH) singularity. The theory with bosons, on the contrary, yields a singular matrix model in the massless case.\footnote{According to the discussion in the previous paragraph, these well-known facts in random matrix theory can be reinterpreted as stemming from Coleman's no-go theorem \cite{Coleman:1973ci} for massless bosons in QCD$_2$.} Here we recognize the singularities encountered in \cite{Russo:2020eif} as the remnants of the IR singularities due to massless fields, and resolve them via mass deformation.\par
				
				\subsection*{On the parameter $\mu $ }
				
				Notice that there is a slight difference in the definition of $\mu$ between the fermionic and the bosonic theory in \eqref{MM0}. In the first case $\mu > 1 $ is real, while in the second case $\mu \in \C$ with $\vert \mu \vert > 1$. It is easy to see that the phase of $\mu \in \C$ can be reabsorbed in a rotation of the integration contour $\ct$, and we henceforth restrict our attention to a real $\mu > 1$ in both cases, with the understanding that for bosons $\mu$ really means $\vert \mu \vert$.\par 
				
				Besides, treating $\mu$ as a real variable with this caveat in mind, the integrand in \eqref{MM0} is analytic in $\mu>1$, and therefore the results for any other $\mu \in \C$ with $\vert \mu \vert >1$ are obtained from analytic continuation of our results. In particular, we could not attain negative values of $\mu$ moving along the real line, because we would cross the FH singularity. Nevertheless, it is possible to take a path from $\mu >1$ to any $\mu^{\prime}<-1$ that runs in the complex plane outside the unit disk.\par
				\begin{rmk}
				The independence of the partition function on $\arg \mu$ is the $U(1)$ freedom to choose the origin of $\ct$, and is the incarnation of the residual diagonal $U(1) \subset U(N)$ gauge symmetry. Our choice $\arg \mu \in 2 \pi \mathbb{Z}$ fixes this residual gauge freedom.
				\end{rmk}

			\subsection*{Notation}
				We introduce the notation
				\begin{equation*}
					Y = \frac{1}{\lambda} , \qquad	\tau = \epsilon \frac{K}{N}
				\end{equation*}
				for, respectively, the inverse of the 't Hooft coupling and a real Veneziano parameter, whose sign carries information on the type of fields we consider.\par
			The parameter space of the theory is 
				\begin{equation*}
					\mathfrak{M} = \left\{ \left( \mu, \tau, Y \right) \in ( 1, \infty ) \times \R^2   \right\} .
				\end{equation*}
				\begin{rmk}
				For the role of the mass as a regulator (for the FH singularities on the mathematical side, for the IR singularities on the field theory side), we do not expect the continuation from $\mathfrak{M}$ to the sheet $\left\{  \mu =1 \right\} \times \R^2 $, studied in \cite{Russo:2020eif,Russo:2020puy}, to be analytic.
				\end{rmk}

		\subsection{Exact finite $N$ evaluations}
		\label{sec:exact}
		
			We now present various analytical results for the matrix model \eqref{MM0}.\par
			It is possible to evaluate the partition function of any unitary matrix model at finite $N$ via the Heine--Szeg\H{o} identity, that for \eqref{MM0} gives 
			\begin{equation}
			\label{eq:Zdet}
				\mz_{U(N)} ^{\epsilon, K} \left( \lambda, \mu \right) = N! \det_{1 \le j,k \le N} \left[  \mathbf{Z}_{jk} \right],
			\end{equation}
			with 
			\begin{equation*}
				  \mathbf{Z}_{jk} =  \sum_{p=0} ^{K }  \frac{K !}{ p! (K -p)! } (- \mu)^{p} (1+\mu^2)^{K -p } \frac{ \dd^p \ }{ \dd x^p } I_{k-j } (2x) \vert_{x= \frac{N}{\lambda} } 
			\end{equation*}
			where $I_k (x)$ is the modified Bessel function. We have simplified the expression assuming $\epsilon=+1$, although a similar determinant expression exists for $\epsilon=-1$ as well. Formula \eqref{eq:Zdet} allows to efficiently compute $\mz_{U(N)} ^{\epsilon, K} $ exactly for fixed $N$ and $K$. This is done in Table \ref{tab:ZexactNK} in Appendix \ref{app:finiteN}.\par
            \medskip
				The partition function for generic masses, encoded in the parameters $(\mu_1, \dots , \mu_K)$, can be also related to the expectation value of Wilson loops in arbitrary representations in the pure GWW model, thanks to the Cauchy identity (see \cite{Santilli:2020snh} for a similar procedure). For the fermionic theory we write 
				\begin{align*}
					\mz _{ U(N) } ^{+1,K} & = \left(  \prod_{f=1} ^{K} \mu_f ^2 \right) \oint_{\ct ^N} \prod_{1 \le j < k \le N } \vert z_j - z_k \vert^2 ~ \prod_{j=1} ^{N} \left[ \prod_{f=1} ^{K} \left( 1- \frac{z_j}{\mu_f } \right) \left( 1 - \frac{ \bar{z}_j }{ \mu_f } \right)  \right] e^{ \frac{N}{\lambda} (z_j + \bar{z}_j ) } ~ \frac{ \dd z_j }{ 2 \pi \ii z_j } \\
					& = \left(  \prod_{f=1} ^{K} \mu_f ^2 \right)  \sum_{ \mathcal{R}_1 } \sum_{ \mathcal{R}_2 } \mathbf{s}_{ \mathcal{R}_1 ^{\prime} } (- \mu_1 ^{-1} , \dots, - \mu_K ^{-1} )  \mathbf{s}_{ \mathcal{R}_2 ^{\prime} } (- \mu_1 ^{-1} , \dots, - \mu_K ^{-1} )    \\ 
					& \times \oint_{\ct ^N} \prod_{1 \le j < k \le N } \vert z_j - z_k \vert^2 ~   \mathbf{s}_{ \mathcal{R}_1 } ( z_1 , \dots , z_N )  \mathbf{s}_{ \mathcal{R}_1 } ( \bar{z}_1 , \dots , \bar{z}_N ) ~ \prod_{j=1} ^{N} e^{ \frac{N}{\lambda} (z_j + \bar{z}_j ) } ~ \frac{ \dd z_j }{ 2 \pi \ii z_j } 
				\end{align*}
				where the sum runs over Young diagrams $\mathcal{R}$ of length at most $N$ and the first row at most $K$, $\mathcal{R}^{\prime}$ is the conjugate diagram, and $ \mathbf{s}_{ \mathcal{R} } $ is the corresponding Schur polynomial, that is, the character of the irreducible representation associated to the diagram $\mathcal{R}$. In the last line we identify the correlator of two Wilson loops in the pure GWW model, 
				\begin{equation}
				\label{ZCauchy}
					\frac{  \mz _{ U(N) } ^{+1,K} }{  \mz_{U(N)} ^{ (\mathrm{GWW}) }  } = \left(  \prod_{f=1} ^{K} \mu_f ^2 \right)  \sum_{ \mathcal{R}_1 } \sum_{ \mathcal{R}_2 } \mathbf{s}_{ \mathcal{R}_1 ^{\prime} } (- \mu_1 ^{-1} , \dots, - \mu_K ^{-1} )  \mathbf{s}_{ \mathcal{R}_2 ^{\prime} } (- \mu_1 ^{-1} , \dots, - \mu_K ^{-1} )   ~ \langle \mathcal{W}_{\mathcal{R}_1}  \overline{\mathcal{W}}_{\mathcal{R}_2}  \rangle^{(\mathrm{GWW})}  ,
				\end{equation}
				with $\overline{\mathcal{W}}$ meaning that the Wilson loop involves conjugated variables. We can in principle further expand the product of the two Schur polynomials with the Littlewood--Richardson rule, but this would entail inverting the variables $\bar{z}_1, \dots , \bar{z}_N$ in the second Schur polynomial, as in \cite{Santilli:2020snh}.\par

				Expression \eqref{ZCauchy} is suggestive but not very useful as it is, because the vacuum expectation value (vev) of a Wilson loop in a generic representation is not known in closed form. Nevertheless, we can go deeper in the character expansion thanks to the formula \cite{Bars}
				\begin{equation*}
					\exp \left(  \frac{N}{\lambda} \tr U  \right) = \sum_{\mathcal{R}} \left( \frac{N}{\lambda} \right)^{\vert \mathcal{R} \vert } ~ \dim \mathcal{R} ~ \left( \prod_{j=1}^{N} \frac{(N-j)!}{(N-j + \mathcal{R}_j)! }  \right) ~ \mathbf{s}_{ \mathcal{R} } ( z_1 , \dots , z_N ) .
				\end{equation*}
				Using this equation and its conjugate and applying twice the Littlewood--Richardson rule we get 
				\begin{align*}
					\mz _{ U(N) } ^{+1,K} & = \left(  \prod_{f=1} ^{K} \mu_f ^2 \right)  \sum_{ \mathcal{R}_1 , \mathcal{R}_2, \mathcal{R}_3, \mathcal{R}_4 } \mathbf{s}_{ \mathcal{R}_1 ^{\prime} } (- \mu_1 ^{-1} , \dots, - \mu_K ^{-1} )  \mathbf{s}_{ \mathcal{R}_2 ^{\prime} } (- \mu_1 ^{-1} , \dots, - \mu_K ^{-1} ) \\
					& \times  \left( \frac{N}{\lambda} \right)^{\vert \mathcal{R}_3 \vert + \vert \mathcal{R}_4 \vert} ~ \dim \mathcal{R}_3  \dim \mathcal{R}_4 ~ \left[ \prod_{j=1}^{N} \frac{\left( (N-j)! \right)^2 }{(N-j + \mathcal{R}_{3,j} )! (N-j + \mathcal{R}_{4,j} )! }  \right] ~   c_{13;24}
				\end{align*}
				where $c_{13;24} \equiv \sum_{ \mathcal{R}} c_{\mathcal{R}_1 \mathcal{R}_3 } ^{\mathcal{R}}  c_{\mathcal{R}_2 \mathcal{R}_4 } ^{\mathcal{R}} $, with $c_{\mathcal{R}_j \mathcal{R}_k} ^{\mathcal{R}}$ the Littlewood--Richardson coefficients, and we have used the orthogonality of the Schur polynomials. The bosonic model $\mz _{ U(N) } ^{-1,K}$ admits a closely related expression, with the partitions $\mathcal{R}_1$ and $\mathcal{R}_2$ instead of their conjugate in the first line and dropping the restriction on the first rows.\par
				Finally, there is an additional, simpler although more formal closed form expression for $\mz_{U(N)} ^{\epsilon, K}$   \cite{Adler99}
				\begin{equation*}
				    \mz_{U(N)} ^{\epsilon, K} = \sum_{\mathcal{R}} \left(    \mathbf{s}_{ \mathcal{R}}  \left(  NY - \epsilon  \frac{K}{\mu} ,  - \epsilon  \frac{K}{\mu^2}, - \epsilon  \frac{K}{\mu^3} , \dots \right) \right)^2
				\end{equation*}
				where now the Schur functions must be interpreted as characters of $U(\infty)$, and the sum runs over Young diagrams $\mathcal{R}$ with at most $N$ rows.

		\section{Phase structure}
		\label{sec:Phases}
			This section is dedicated to the large $N$ analysis of the matrix model \eqref{MM0} and the determination of its phase diagram.\par
			Write the partition function \eqref{MM0} as 
				\begin{equation}
				\label{eq:MM1}
					\mz_{U(N)}  = \oint_{\ct^{N}} e^{ - N^2 S_{\mathrm{eff}} \left( z_1, \dots, z_N \right) }~ \prod_{j=1} ^{N} \frac{ \dd z_j}{ 2 \pi \ii z_j} 
				\end{equation}
				where the effective action $S_{\mathrm{eff}} $ is the sum of a potential and the Coulomb interaction between eigenvalues:
				\begin{align*}
					S_{\mathrm{eff}} \left( z_1, \dots, z_N \right) & = \frac{1}{N} \sum_{j=1} ^{N} V_{\mathrm{eff}} (z_j) +  \frac{1}{N^2} \sum_{j=1} ^{N} \sum_{k \ne j } V_{\mathrm{int}} \left( z_j , z_k  \right) \\
					V_{\mathrm{eff}} \left( e^{\ii \theta} \right) &= - \frac{2}{\lambda} \cos \theta - \tau \log \left( 1 + \mu^2 - 2 \mu \cos \theta  \right) \\
					V_{\mathrm{int}} \left( e^{\ii \theta} , e^{\ii \varphi}  \right) & = - \log 2 \sin \left( \frac{ \theta - \varphi }{2}  \right) .
				\end{align*}
				writing $z \in \ct$ as $z= e^{\ii \theta }$, $- \pi < \theta \le \pi$.\par
				The potential $V_{\mathrm{eff}} \left( e^{\ii \theta} \right)$ admits isolated minima at each point in $\mathfrak{M}$. Besides, there exists a surface $\left\{ \lambda = \lambda_{\ast} (\mu, \lambda )  \right\} \subset \mathfrak{M}$ at which it passes from a single-well to a double-well profile. Explicitly, these two regimes are separated by 
				\begin{equation*}
					\lambda_{\ast} (\mu, \tau) =   \frac{(\mu -1)^2 }{\tau \mu}  ,
				\end{equation*}
				and the potential develops stationary points at $\theta = \pm \theta_{\ast}$ with 
				\begin{equation*}
					\tan \theta_{\ast} = \pm \frac{\sqrt{\mu ^2 \left(2-\lambda ^2 \tau ^2\right)+2 \lambda  \mu ^3 \tau +2 \lambda \mu  \tau -\mu ^4-1}}{-\lambda  \mu  \tau +\mu ^2+1} .
				\end{equation*}
				We stress that $\lambda_{\ast}$ is not a critical value of the model \eqref{MM0}.\par
\begin{figure}[htb]
			\centering
				\includegraphics[width=0.35\textwidth]{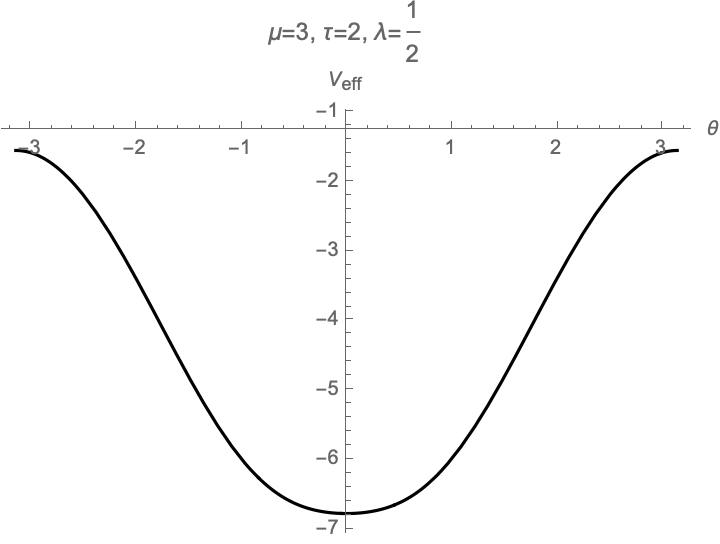}
				\hspace{0.05\textwidth}
				\includegraphics[width=0.35\textwidth]{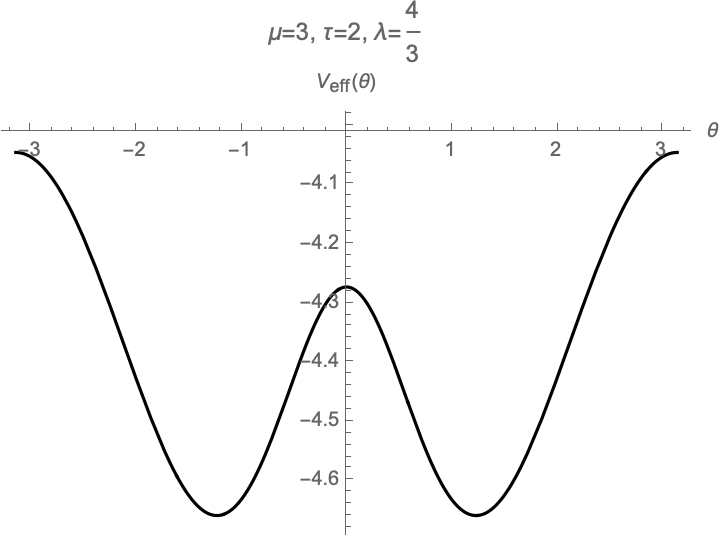}
				\caption{$V_{\mathrm{eff}} (e^{\ii \theta})$ at $\mu=3$ and $\tau=2$. Left: $\lambda= \frac{1}{2}$. Right: $\lambda= \frac{4}{3}$.}
			\label{fig:Veffpos}
			\end{figure}
\begin{figure}[htb]
			\centering
				\includegraphics[width=0.35\textwidth]{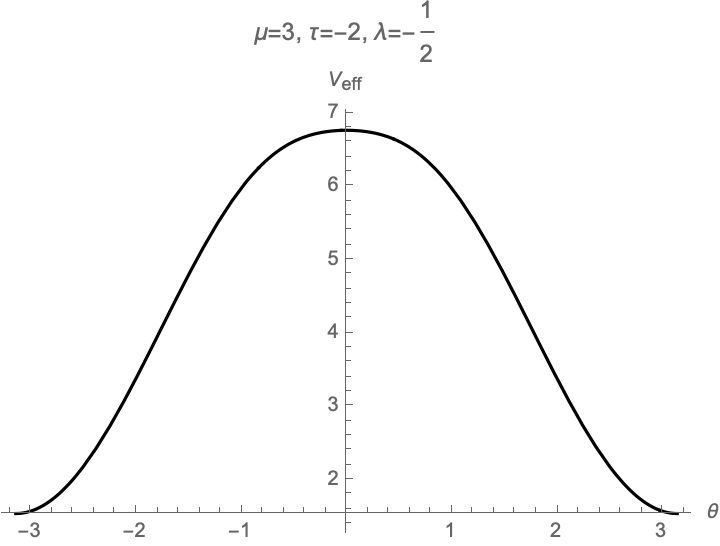}
				\hspace{0.05\textwidth}
				\includegraphics[width=0.35\textwidth]{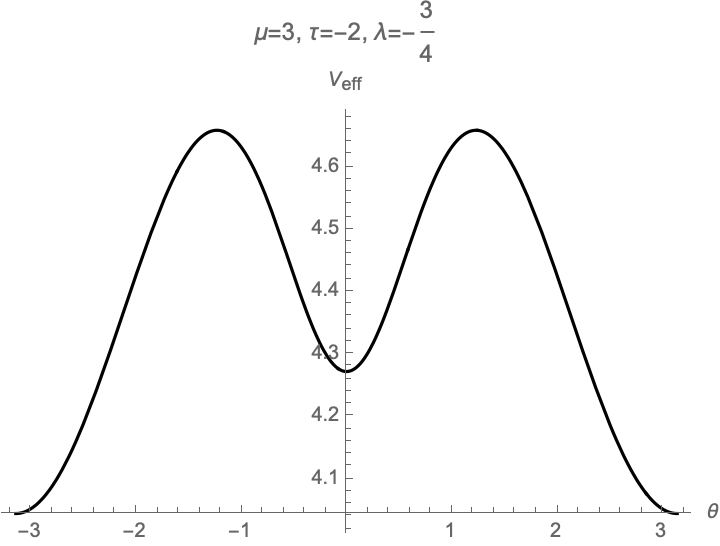}
				\caption{$V_{\mathrm{eff}} (e^{\ii \theta})$ at $\mu=3$ and $\tau=-2$. Left: $\lambda=-\frac{1}{2}$. Right: $\lambda=-\frac{4}{3}$.}
			\label{fig:Veffneg}
			\end{figure}

				The potential is plotted in Figure \ref{fig:Veffpos} ($\tau >0$) and in Figure \ref{fig:Veffneg} ($\tau <0$). Clearly, the two figures have the same shape but upside down. Nonetheless, it is important to distinguish the role of minima and maxima to understand the phase structure.\par
			
			\par
			\par
			\medskip
				Now that we have set the ground, we are ready to discuss the large $N$ limit of the model \eqref{MM0}.

				\subsection{Large $N$}
				\label{sec:LargeNphases}
			
				We now take the large $N$ 't Hooft and Veneziano limit of the partition function \eqref{eq:MM1}. This means that we consider the planar limit with both $\lambda$ and $\tau$ fixed. The leading contributions to the integral at large $N$ come from the saddle points of the effective action:
				\begin{equation}
				\label{finiteNSPE}
					\frac{ \partial S_{\mathrm{eff}} }{ \partial \theta_j } = 0 \qquad j=1, \dots, N .
				\end{equation}
				Introducing the eigenvalue density 
				\begin{equation*}
					\rho (\theta) = \frac{ 2 \pi }{N} \sum_{j=1} ^{N}  \delta \left( e^{\ii \theta} - e^{\ii \theta_j} \right) ,
				\end{equation*}
				with normalization chosen so that 
				\begin{equation}
				\label{normrho}
					\int_{- \pi} ^{\pi} \frac{\dd \theta }{ 2 \pi } \rho (\theta ) = 1 ,
				\end{equation}
				we can collect the system \eqref{finiteNSPE} of $N$ coupled equations in a single singular integral equation at large $N$. The saddle point equation then reads 
				\begin{equation}
				\label{SPE1}
					\mathrm{P} \int_{- \pi } ^{\pi}  \frac{\dd \varphi }{ 2 \pi } \rho (\varphi )  ~ \cot \left( \frac{ \theta - \varphi}{2}  \right) = 2 Y \sin \theta - \frac{  2 \mu \tau \sin \theta }{ 1+ \mu^2 - 2 \mu \cos \theta } .
				\end{equation}
				The solution to \eqref{SPE1} must satisfy the non-negativity constraint 
				\begin{equation}
				\label{rhoge0}
					\rho (\theta) \ge 0 , \qquad - \pi < \theta \le \pi ,
				\end{equation}
				that follows from the compactness of the integration domain.\par

					\subsection*{Ungapped solution: Phase 0}
						We begin assuming $\rho (\theta)$ is supported on the whole circle, $- \pi < \theta \le \pi$. We exploit $\mu >1$ to obtain the solution \cite{Mandal:1989ry}
						\begin{equation}
						\label{eq:rho0}
							\rho_{0} (\theta) = 1 + 2 Y  \cos \theta - 2 \tau \frac{ \mu \cos \theta -1 }{ 1 + \mu^2 - 2 \mu \cos \theta }  .
						\end{equation}
						The derivation is standard \cite{Gross:1980he,Santilli:2020ueh}, thus we omit it. We call Phase 0 the region of $\mathfrak{M}$ for which the solution \eqref{eq:rho0} is valid.

					\subsection*{Critical loci}
						In those regions of $\mathfrak{M}$ for which the solution \eqref{eq:rho0} violates the constraint \eqref{rhoge0}, we should drop the assumption $\mathrm{supp} \rho = (- \pi, \pi]$ and look for a new solution, whose support has one or more gaps on the unit circle. The arcs on which $\rho$ is supported are called cuts.\par
\begin{figure}[htb]
							\centering
								\includegraphics[width=0.31\textwidth]{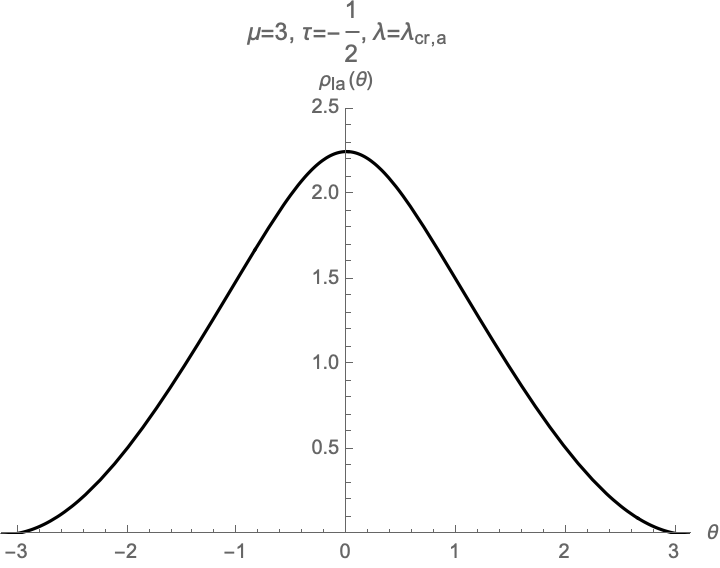}
								\hspace{0.01\textwidth}
								\includegraphics[width=0.31\textwidth]{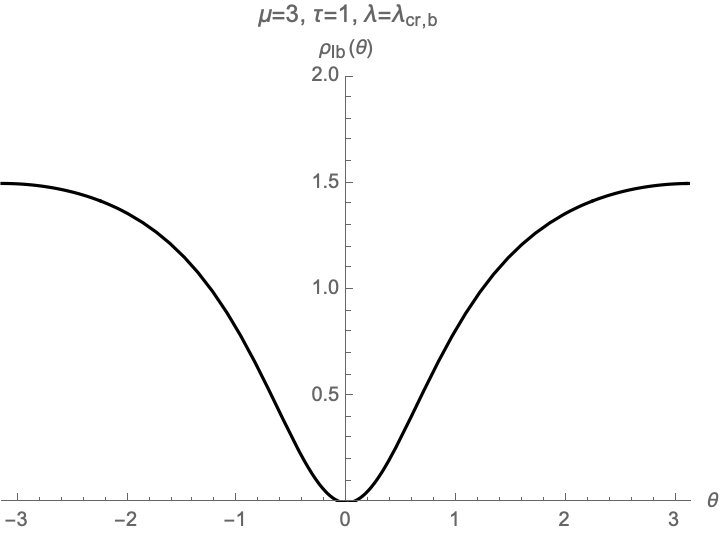}
								\hspace{0.01\textwidth}
								\includegraphics[width=0.31\textwidth]{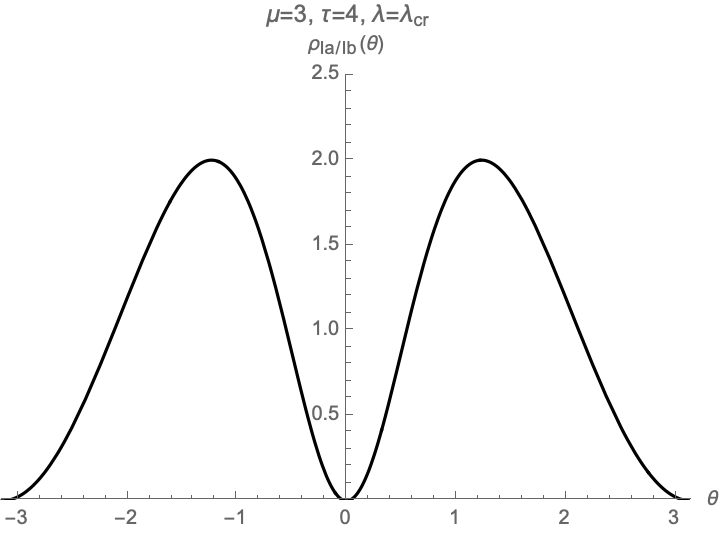}
								\caption{$\rho_0 (\theta)$ at the transition point. Left: $\mu=3$, $\tau=- \frac{1}{2}$ and $\lambda =\lambda_{\mathrm{cr,a}}$. Centre: $\mu=3$, $\tau=1$ and $\lambda = \lambda_{\mathrm{cr,b}}$. Right: $\mu=3$, $\tau=4$ and $\lambda= \frac{2}{3}$.}
							\label{fig:rhocritAandB}
						\end{figure}

						We find a phase transition with a gap opening at $\theta = \pm \pi$ at the critical surface 
						 \begin{equation}
						 \label{Lambdacr1}
			 				Y_{\mathrm{cr,a}} = \frac{1}{2}  + \frac{ \tau}{\mu + 1} .
						 \end{equation}
						 Another phase transition, with a gap opening at $\theta=0$, takes place at the critical surface 
						 \begin{equation}
						 \label{Lambdacr2}
				 			Y_{\mathrm{cr,b}}  = - \frac{1}{2}  + \frac{ \tau}{\mu - 1}  .
						 \end{equation}
						 Besides, there exists a multi-critical point at the value $\tau = \tau_{\mathrm{cr},+} (\mu)$ at which $\rho_0 (\pm \pi) = 0 = \rho_0 (0)$, determined as the unique point at which $Y_{\mathrm{cr,a}} $ and $Y_{\mathrm{cr,b}} $ meet: 
						 \begin{equation}
						 	\tau_{\mathrm{cr},+} (\mu ) = \frac{ \mu^2 -1 }{2} .
						 \end{equation}
						Examples of the limiting cases of $\rho_0 (\theta)$ are shown in Figure \ref{fig:rhocritAandB}.\par
						 	\par
\begin{figure}[htb]
						\centering
							\includegraphics[width=0.4\textwidth]{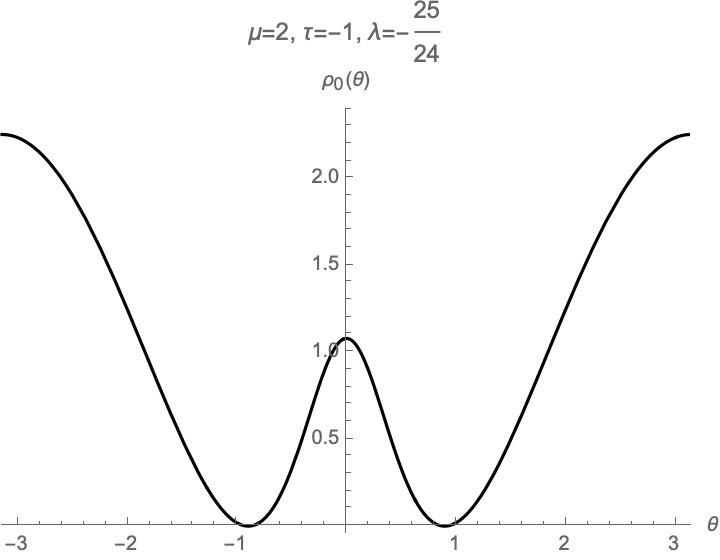}
							\caption{$\rho_0 (\theta)$ at the transition point. $\mu=2$, $\tau=-1$ and $\lambda=-\frac{25}{24}$. The support of $\rho$ will break in two disjoint cuts beyond this critical value.}
						\label{fig:rhocritC}
						\end{figure}

						Besides the two critical surfaces just described, looking at $\rho_0 (\theta)$ for negative $Y$ and $\tau$ we also find values at which it attains zero value at two distinct, symmetric points in the interior of $(-\pi, \pi)$, as in Figure \ref{fig:rhocritC}. We expect a new phase transition into a two-cut solution.
						\par

			 		\subsection*{One-cut solution: Phase Ia}
			 			 \label{sec:PhaseI}
			 
			 		We now solve Equation \eqref{SPE1} dropping the assumption that $\rho (\theta)$ is supported on the whole $\ct$, and replace it by the assumption that the support is an arc $\Gamma \subset \ct$. The derivation is standard and we relegate it to Appendix \ref{app:derivomega}.\par
			 		Introduce the trace of the resolvent in the large $N$ limit, 
			 		\begin{equation}
			 		\label{def:resolvent}
			 			\omega (z) = \int_{\Gamma} \frac{ \dd w}{2 \pi \ii w} \varrho (w) \frac{ z+w}{z-w} , \qquad z \in \mathbb{C} \setminus \Gamma .
			 		\end{equation}
			 		We adopt the standard notation 
			 		\begin{equation*}
			 			\omega_{\pm} (e^{\ii \theta}) \equiv  \lim_{\epsilon \to 0^{+}} \omega (z = (1 \pm \epsilon)e^{\ii \theta}) .
			 		\end{equation*}
			 		Then 
			 		\begin{equation*}
			 			\omega_{+} (e^{\ii \theta}) - \omega_{-} (e^{\ii \theta}) = 2 \varrho (e^{\ii \theta}) , \qquad e^{\ii \theta} \in \Gamma .
			 		\end{equation*}
			 		We find (see Appendix \ref{app:derivomega} for the details)
			 		\begin{equation*}
			 			\omega_{\mathrm{Ia}} (z) = - \ii W (z) + \sqrt{\left(e^{\ii \theta_0} - z \right)  \left(e^{- \ii \theta_0} - z \right) } \left[ Y  \left( 1+ \frac{1}{z} \right) - \frac{\tau }{ \sqrt{1 + \mu^2 - 2 \mu \cos \theta_0} } \left( \frac{\mu }{z- \mu}  -  \frac{1}{z- \mu^{-1}}  \right) \right] .
			 		\end{equation*}
			 		The first term is regular and, taking the discontinuity at $z= e^{\ii \theta } \in \Gamma $, we arrive at  
			 		\begin{equation}
			 		\label{rhoIa}
			 			\rho_{\mathrm{Ia}} (\theta ) = 2 \cos \frac{\theta}{2} \cdot \sqrt{ 2 \cos \theta - 2 \cos \theta_0 } \cdot \left[ Y  - \frac{\tau \mu (\mu-1) }{\sqrt{1 + \mu^2 - 2 \mu \cos \theta_0 } \left( 1 + \mu^2 - 2 \mu \cos \theta \right)  }  \right]
			 		\end{equation}
			 		The angle $\theta_0$ is fixed by normalization: 
			 		\begin{equation}
			 		\label{eq:condy0Ia}
			 			Y \left( 1-y_0 \right) + \tau \left( \frac{ \mu -1}{ \sqrt{ 1 + \mu^2 - 2 \mu y_0} } -1  \right) = 1 .
			 		\end{equation}
			 		where $y_0 := \cos \theta_0$. Equation \eqref{eq:condy0Ia} admits a unique real solution, thus the problem is completely determined.

					 \subsection*{One-cut solution: Phase Ib}

			 		The solution above has been derived assuming that $\Gamma$ is an arc along $\ct$ joining $e^{- \ii \theta_0}$ to $e^{\ii \theta_0}$ running counter-clockwise, thus the gap has opened around $\theta=\pi$. For the gap opening at $\theta=0$, the procedure is identical, but now $\Gamma$ is an arc from $\tilde{\theta}_0 >0$ to $2 \pi - \tilde{\theta}_0$. The procedure of Appendix \ref{app:derivomega} leads us to 
			 		\begin{equation*}
			 			\rho_{\mathrm{Ib}} (\theta ) =  2 \left\lvert  \sin \frac{\theta}{2} \right\rvert  \sqrt{ 2 \cos \tilde{\theta}_0 - 2 \cos \theta } \left[ - Y  + \frac{\tau \mu (\mu+1) }{\sqrt{1 + \mu^2 - 2 \mu \cos \tilde{\theta}_0 } \left(  \mu^2 +1 -2 \mu \cos \theta  \right)  }  \right]
			 		\end{equation*}
			 		which is non-negative definite. There is, however, a more direct route to get the correct answer. Looking back at the matrix model \eqref{MM0} we can chose a different parametrization $0 \le \theta < 2 \pi$, and the solution with the gap opening at $\theta=0$ is recovered from the solution \eqref{rhoIa} in Phase Ia upon replacement $Y \mapsto - Y$, $\mu \mapsto - \mu$ and eventually $\theta + \pi \mapsto \theta$.\par
			 		In conclusion, we have two different phases with a one-cut solution, as expected: one for $Y > Y_{\mathrm{cr,a}} (\tau, \mu) $, that we have called Phase Ia, and one for $Y < Y_{\mathrm{cr,b}} (\tau, \mu)$, that we have called Phase Ib.

			 		\subsection*{Two-cut solution: Phase II}
			 		\label{sec:PhaseII}

\begin{figure}[htb]
			 		\centering
			 			\includegraphics[width=0.4\textwidth]{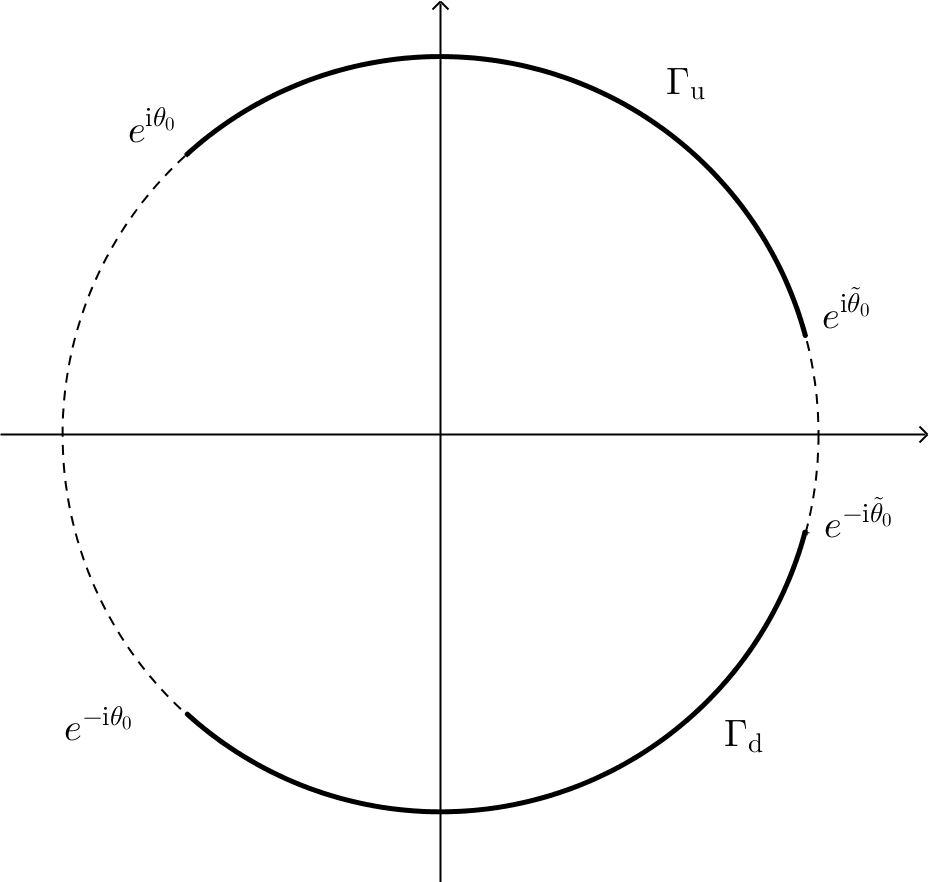}
			 		\caption{The two-cut support $\Gamma$ in Phase II.}
			 		\label{fig:Gammatwo}
			 		\end{figure}

			 			We have seen that at $\tau = \tau_{\mathrm{cr},+} =( \mu^{2}-1)/2$ the critical surfaces $Y = Y_{\mathrm{cr,a}}$ and $Y=Y_{\mathrm{cr,b}}$ meet. Thus, we expect a new phase characterized by a two-cut solution in the region 
			 			\begin{equation*}
			 				\left\{  (\mu, \tau, Y) \ : \ \mu>1, \tau> \frac{\mu^{2}-1}{2} , Y_{\mathrm{cr,a}} < Y < Y_{\mathrm{cr,b}} \right\} \subset \mathfrak{M} .
			 			\end{equation*}
			 			with gaps around $\theta=0$ and $\theta=\pm \pi$, and eigenvalue density supported on 
			 			\begin{equation*}
			 				\mathrm{supp} \rho_{\mathrm{II}} = \Gamma \cong  \Gamma_{\mathrm{u}} \sqcup \Gamma_{\mathrm{d}} := \left\{  e^{\ii \varphi } \in \ct \ : \ \tilde{\theta}_0 \le \theta \le \theta_0  \right\} \sqcup \left\{  e^{\ii \varphi } \in \ct \ : \ - \theta_0 \le \theta \le  - \tilde{\theta}_0   \right\} .
			 			\end{equation*}
			 			That is, $\Gamma$ is the union of two disjoint arcs, $\Gamma_{\mathrm{u}}$ and $\Gamma_{\mathrm{d}}$, as in Figure \ref{fig:Gammatwo}.\par
			 			
			 		\par
			 			
			 			To determine $\rho_{\mathrm{II}} (\theta)$ it is simpler to adopt a different strategy, detailed in Section \ref{sec:stereo} below.

			 	\subsection*{Two-cut solution: Phase III}
			 		\label{sec:PhaseIII}
			 			The fact that the potential $V_{\mathrm{eff}} (e^{\ii \theta})$ develops a double well for negative $Y$ and $\tau$ in a given range hints at the existence of a two-cut solution in that region of $\mathfrak{M}$, with the eigenvalues sitting around the two minima. This observation is corroborated looking at the shape of $\rho_{\mathrm{Ia}} (\theta)$ and $\rho_{\mathrm{Ib}} (\theta)$ in the negative quadrant, where they become negative in $Y_{\mathrm{cr,b}} < Y < Y_{\mathrm{cr,a}}$ for $\tau $ below a certain threshold.\par
			 			We find a transition from Phase 0 to a two-cut phase in 
			 				\begin{equation*}
			 					Y_{\mathrm{cr,c}+} < Y < Y_{\mathrm{cr,a}} \quad \text{ and } \quad Y_{\mathrm{cr,b}} < Y < Y_{\mathrm{cr,c}-} 
			 				\end{equation*}
			 				where the critical surfaces $Y = Y_{\mathrm{cr,c}\pm}$ are given by 
			 				\begin{equation*}
			 					Y_{\mathrm{cr,c} \pm } (\tau, \mu) = \frac{ \mu }{(\mu^2+1)^2} \left[   \mu^2 (\tau -1) - 3 \tau - 1 \pm 2 \sqrt{ - \tau \left[ 2 \tau (\mu^2 -1) + (\mu^4 -1)  \right] }   \right] .
			 				\end{equation*}
			 				The two curves $Y_{\mathrm{cr,\pm}}$ form an ellipse in each $(\tau, Y)$-leaf of $\mathfrak{M}$ at fixed $\mu$, with the physical critical curve being the first branch of the ellipse encountered when decreasing $Y$ from 0.\par
\begin{figure}[htb]
			 			\centering
			 				\includegraphics[width=0.4\textwidth]{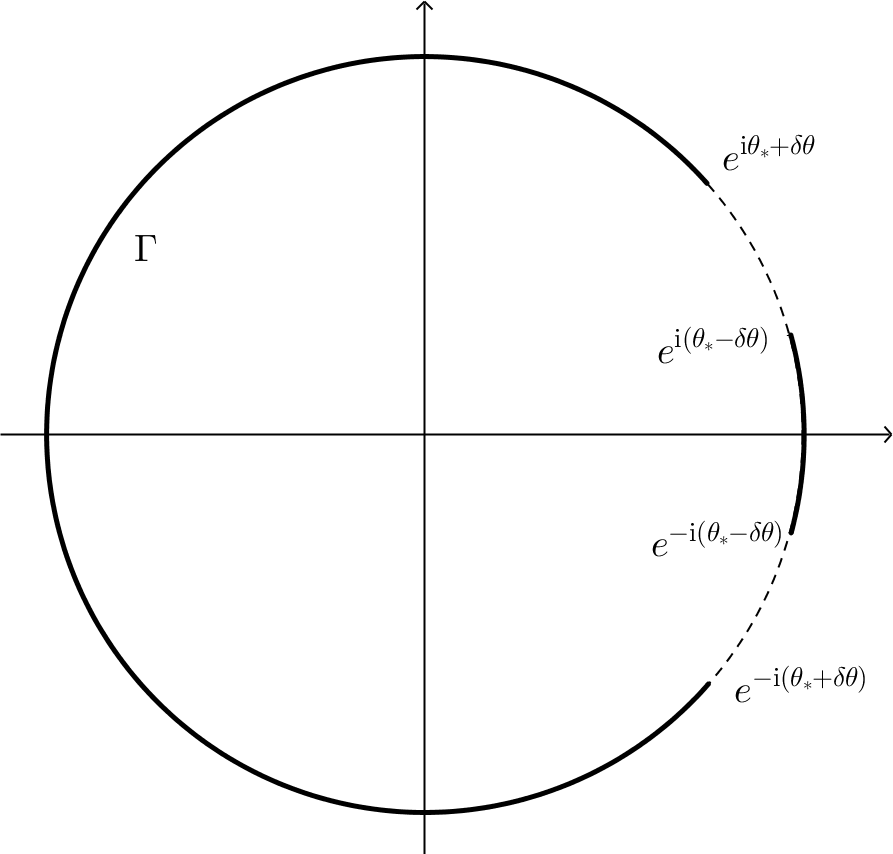}
			 			\caption{The two-cut support $\Gamma $ in Phase III.}
			 			\label{fig:GammaPhIII}
			 			\end{figure}

			 				In this phase, that we call Phase III, the eigenvalues distribute along a contour $\Gamma$ which consists of two cuts, with gaps opening around $\pm \theta_{\ast}$, see Figure \ref{fig:GammaPhIII}.

			 			\par
			 					The eigenvalue density is 
			 					\begin{align}
			 						\rho_{\mathrm{III}} (\theta) & =  2 \sqrt{ \left[ \cos \left( \theta_{\ast} - \delta \theta \right) - \cos \theta \right]   \left[ \cos \left( \theta_{\ast} + \delta \theta \right) - \cos \theta \right] } \notag \\
			 						& \times \left[ - Y + \frac{ \tau \mu (\mu+1)(\mu-1) }{ \sqrt{ \left( \mu^2 +1 - 2 \mu \cos  \left( \theta_{\ast} - \delta \theta \right)   \right) \left(  \mu^2 +1 - 2 \mu \cos  \left( \theta_{\ast} + \delta \theta \right)   \right)  } \left[\mu^2 +1 - 2 \mu \cos \theta  \right] }  \right] .
			 					\end{align}
			 					Note that the argument of the outer square root is non-negative definite. The value of $\theta_{\ast}$ is known explicitly, as obtained from Phase 0, and the dependence of $\delta \theta$ on the parameters is fixed by normalization. Equivalently, we can fix $\cos \left( \theta_{\ast} + \delta \theta \right) $ and $\cos \left( \theta_{\ast} - \delta \theta \right) $ comparing the large $z$ behaviour of $\omega (z)$ computed in this phase with its definition.\par
			 				\medskip
			 				For multi-cut solutions, the dependence on the number of eigenvalues filling each cut should be taken into account when computing physical observables \cite{Bonnet:2000dz}. We analyze the role of the filling fractions in Appendix \ref{app:fillingfrac}: the upshot is that our conclusions are unaltered, both in phase II and III, although for different reasons.

			 			\subsection{Phase diagram}
			 			
\begin{figure}[htb]
			 			\centering
			 				\includegraphics[width=0.7\textwidth]{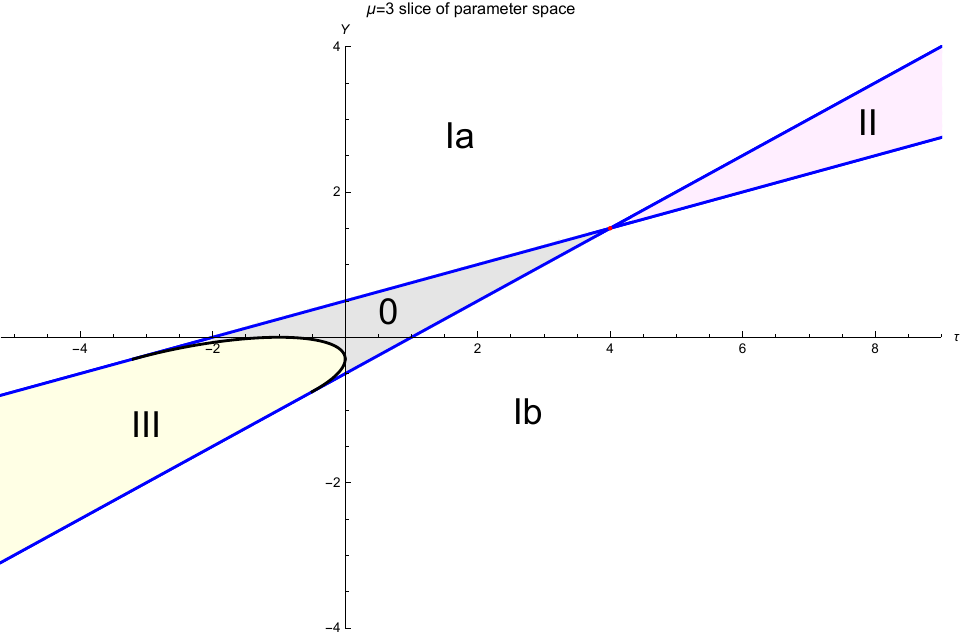}
			 			\caption{Phase diagram of the model in the $(\tau, Y)$ plane, at $\mu=3$. The blue straight lines are $Y = Y_{\mathrm{cr,a}}$ and $Y = Y_{\mathrm{cr,b}}$, the black curve is $Y = Y_{\mathrm{cr,c}\pm}$, the red dot is the multi-critical point at $ \tau = \frac{ \mu^2-1}{2}$. The gray shaded region in the ungapped phase, Phase 0. The other light shaded regions are the two-cut phases, Phase II and III.}
			 			\label{fig:Phasediag}
			 			\end{figure}

			 			Putting all the information together, the following phase diagram emerges.
			 			\begin{itemize}
			 				\item [0)] When both $Y$ and $\tau$ are small, Phase 0 holds, with the eigenvalues spread on the whole circle.
			 				\item [Ia)] When $Y > Y_{\mathrm{cr,a}}$ the system is in a new phase, Phase Ia, with a one-cut solution gapped around $\theta = \pm \pi$.
			 				\item [Ib)] Likewise when $Y < Y_{\mathrm{cr,b}}$ the system is in Phase Ib, with a one-cut solution gapped around $\theta = 0$.
			 				\item [II)] At $\tau > \frac{ \mu^2 -1}{2}$ the two critical surfaces cross each other. In the region $Y_{\mathrm{cr,a}}< Y < Y_{\mathrm{cr,b}}$ the system is in Phase II, a two-cut solution with density of eigenvalues gapped both around $\theta=0$ and $\theta = \pi$.
			 				\item [III)] The system develops a new two-cut phase, Phase III, in the region $Y_{\mathrm{cr,b}} < Y <Y_{\mathrm{cr,a}} $ and also bounded by an arc of ellipse determined by $Y_{\mathrm{cr,c} \pm}$. The density of eigenvalues is gapped around $\theta = \pm \theta_{\ast}$, with $\theta_{\ast} \to \pi$ as $Y \to Y_{\mathrm{cr,a}}$ and $\theta_{\ast} \to 0$ as $Y \to Y_{\mathrm{cr,b}}$.
			 			\end{itemize}
			 			See Figure \ref{fig:Phasediag} for a slice of $\mathfrak{M}$ at fixed $\mu$.\par
			 			
			 			\par
			 			\medskip
			 			Taking the massless limit $\mu \to 1^{+}$, the critical surface $Y_{\mathrm{cr,b}}$ is rotated onto the vertical axis. Using the analytic dependence on $\mu$, we can also reach $\mu \to -1^{-}$ by first going to the negative real axis walking through $\C$ outside of the unit disk and then taking the limit $\vert \mu \vert \to 1^{+}$. In that case, it is $Y_{\mathrm{cr,a}}$ that is rotated onto the vertical axis.
			 			
			 			\subsection{Free energy and massless theory}
			 			Before delving in the analysis of Wilson loop vevs in the next section, we comment on the free energy of the model, defined as 
			 			\begin{equation*}
			 				\mathcal{F} = \frac{1}{N^2} \log \mathcal{Z}  .
			 			\end{equation*}
			 			The free energy in Phase 0 is easily obtained, and corresponds to the analytic continuation of Szeg\H{o}'s strong limit theorem in the bulk of the 't Hooft parameter space \cite{Santilli:2020trf}. It takes the value 
						\begin{equation}
						\label{FE0}
							\mf_{0} =  Y^2 - 2 Y \frac{\tau }{ \mu}  - \tau^2 \log \left( 1- \frac{1}{\mu^2} \right)  .
						\end{equation}
						It is clearly separated into three contributions: pure gauge ($Y^2$), matter only ($\propto \tau^2$) and the interaction. At strong coupling $\lambda \to \infty$ ($Y \to 0$) we are left with a matter contribution which counts gauge singlets: indeed, the integral over the gauge group projects onto gauge invariant states.\par

                        \subsection*{Massless theory}			
						As we have stressed, a core assumption of our analysis is $\vert \mu \vert >1$, and the massless limit $\vert \mu \vert \to 1^{+}$ can only be taken at the end. Due to the non-analyticity for $\mu \in \ct $, the resulting model will differ from a model with massless matter \cite{Russo:2020eif,Russo:2020puy}.\par
						A main consequence of this non-analyticity is the spontaneous chiral symmetry breaking, that we will discuss in Section \ref{sec:chiralSB}. On the other hand, it is well known that the large $N$ limit and the massless limit do not commute.\par
						In the na\"{i}ve $\vert \mu \vert \to 1$ limit, the free energy in \eqref{FE0} has a logarithmic divergence in the matter contribution. Setting instead $\vert \mu \vert =1$ from the beginning, and $\arg \mu = \tilde{\theta}$, $0 < \tilde{\theta} \le 2\pi $, the partition function acquires a FH singularity and the large $N$ limit cannot be understood by standard methods. We use known results on Toeplitz determinants to derive the free energy in Phase 0 in the massless theory \cite{Widom}: 
						\begin{equation*}
							\mf_{0} \left( \mu = e^{\ii \tilde{\theta}} \right) = Y^2 - 2 Y \tau \cos \tilde{\theta} - 2 \tau^2 \log \left\lvert 2 \sin \frac{\tilde{\theta}}{2} \right\rvert +  \tau^2 \log N .
						\end{equation*}
						That is, the contribution from matter fields has an additional factor of $\log N$ and dominates at large $N$. Remarkably, this matches the logarithmic divergence of the na\"{i}ve massless limit of \eqref{FE0}. The result is in fact much more general \cite{Widom} and directly extends to the case of various Veneziano parameters $\tau_1, \dots, \tau_n$ associated to different $\mu_1, \dots, \mu_n$ that approach the unit circle from outside at different angles $\tilde{\theta}_1 , \dots, \tilde{\theta}_n$.

					\subsection{Stereographic projection}
					\label{sec:stereo}
					
						To better understand Phase II and the transition from a one-cut to a two-cut phase, we map the model onto the real line and study the resulting Hermitian matrix model at large $N$. It can be interpreted as a massive deformation of the model in \cite{Russo:2020pnv}.\par
\begin{figure}[htb]
			 			\centering
			 				\includegraphics[width=0.7\textwidth]{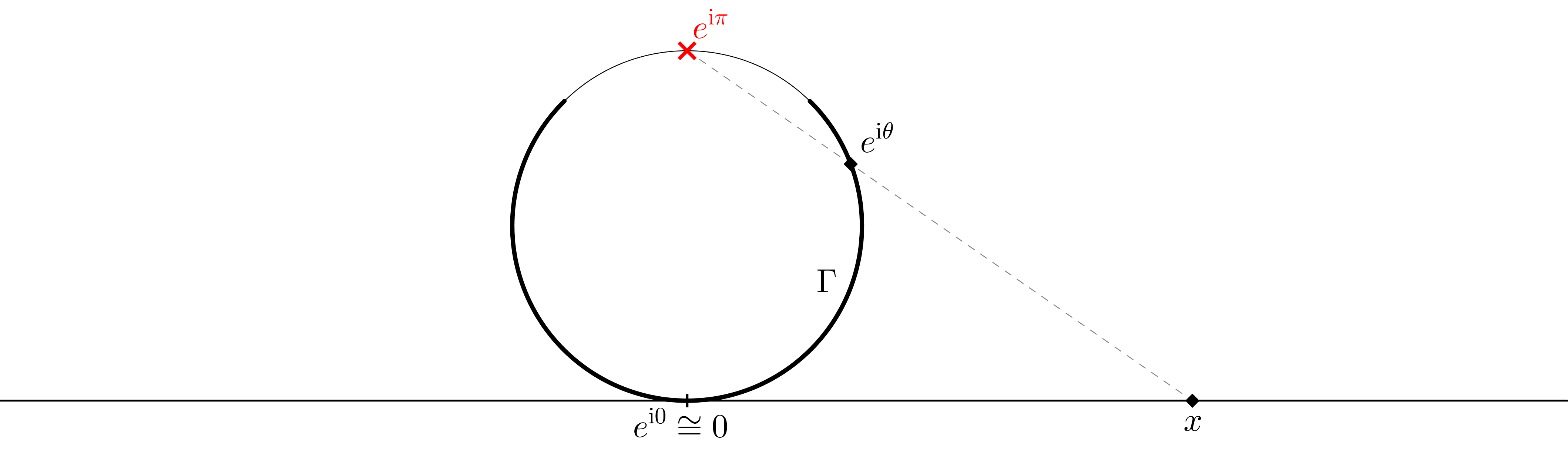}
			 			\caption{The stereographic projection. The red cross is the puncture on the circle, the ticker line is the cut $\Gamma$.}
			 			\label{fig:stereo}
			 			\end{figure}

						We conformally map the unit circle on the real line through the stereographic projection, see Figure \ref{fig:stereo}. The drawback of the stereographic map is that it introduces a puncture on the circle at $\theta = \pm \pi$: this has no effect at finite $N$, but the Hermitian matrix model will fail to reproduce Phase 0 of the unitary matrix model because of this change in topology \cite{Santilli:2020ueh}. Phase 0 and its associated transitions are well understood from the unitary matrix model side, and we use the conformally mapped model as yet another way to gain further insight into the one-cut to two-cut transition.\par
						\par
			 			
			 			Our choice of coordinates is consistent with Phase Ia on the circle, but Phase Ib is easily retrieved rotating $\ct$ by $e^{\ii \pi}$, so that the puncture is placed at $\theta=0$. The Hermitian matrix model is
			 	\begin{equation}
			 			\label{MMstereo}
			 				\mz_{U(N)} ^{\mathrm{p.}}  = \left( \mu -1 \right)^{2K }  \int_{\R^N} \prod_{1 \le j < k \le N} \left( x_j - x_k \right)^2 ~ \prod_{j=1} ^{N} \frac{ ( 1 + \eta^2  x_j ^2 )^{K } }{  (1+x_j ^2)^{K+N  } } ~ e^{2 N Y \left( \frac{1-x_j ^2}{ 1+x_j^2 } \right) } ~  \frac{ \dd x_j }{ 2 \pi}
			 			\end{equation}
			 			with the superscript $\mathrm{p.}$ as notation to remind that it comes from the projection of the model \eqref{MM0}. We have adopted the shorthand notation 
			 			\begin{equation*}
			 				\eta := \frac{ \mu +1}{ \mu -1} , \qquad 1 < \eta < \infty .
			 			\end{equation*}
			 			\begin{rmk}
			 			    Thanks to the mapping of the Vandermonde determinant form the unit circle to the real line that shifts $\tau \mapsto \tau+1$ in the denominator of the integrand in \eqref{MMstereo}, the stability issues pointed out in \cite{Russo:2020pnv} do not arise here. We can thus safely allow $\tau <0$ without spoiling the convergence of the matrix model.
			 			\end{rmk}\par

			 \subsection*{Phase I}

			 			The saddle point equation for the Hermitian matrix model \eqref{MMstereo} is 
			 			\begin{equation*}
			 				\mathrm{P} \int \frac{ \dd y }{ 2 \pi } \frac{  \rho^{\mathrm{p.}} (y) }{ x- y } = x \left[ \frac{4 Y}{\left(x^2+1\right)^2}+\frac{\tau +1}{1+x^2}  -  \frac{\eta^2  \tau }{1 + \eta^2  x^2}  \right] .
			 			\end{equation*}
			 			The solution is found by standard large $N$ techniques \cite{DiFrancesco:1993cyw}. Using a one-cut ansatz for the density $ \rho^{\mathrm{p.}}  (x)$ supported on $[-A,A] \subset \R$ we find 
			 			\begin{equation}
			 				 \rho^{\mathrm{p.}} _{\mathrm{I}} (x)  =  2  \sqrt{ A^2 - x^2 } \left[ \frac{ \tau +1}{ \sqrt{1+A^2}  (1+x^2) }  - \frac{\tau \eta^2 }{\sqrt{1+ \eta^2  A^2 } (1 + \eta^2 x^2 )}  - 2 Y \frac{ A^2 \left( x^2 -1 \right) -2 }{ (1+A^2)^{\frac{3}{2} } (1+x^2)^2 }  \right] .
			 			\end{equation}
			 			The value of $A$ is fixed by normalization: 
			 			\begin{equation*}
			 				\int_{-A} ^{A} \frac{ \dd x }{2 \pi } \rho^{\mathrm{p.}} _{\mathrm{I}} (x)  =  1 \quad \Longrightarrow \quad  \frac{\tau +1}{\sqrt{1+A^2}} - \frac{\tau}{\sqrt{1+\eta^2 A^2}}  + 2 Y \frac{ A^2}{(1+A^2)^{\frac{3}{2}}} =0 .
			 			\end{equation*}
			 			As a cross-check, turning off the mass deformation, $\mu \to 1$, sends $\eta \to \infty$ and we recover the eigenvalue density found in \cite{Russo:2020pnv}. Besides, sending $A^2 \to \infty$ and expanding at leading order in $\frac{1}{A}$ the normalization becomes the consistency condition 
			 			\begin{equation*}
			 				Y = \frac{1}{2} + \frac{ \tau}{ \mu +1} ,
			 			\end{equation*}
			 			correctly reproducing the critical surface $Y_{\mathrm{cr,a}}$ in the limit in which $A$ is back-projected to $e^{\ii \pi}$. We stress that, requiring that $\rho^{\mathrm{p.}} (x) \dd x$ descends from a measure on $\ct$, the non-negativity constraint $\rho (x) \ge 0$ must be imposed.\par
			 			Looking at $\rho^{\mathrm{p.}} _{\mathrm{I}} (0)$, we find that the critical point is fixed by the condition 
			 			\begin{equation*}
			 				2 Y \left( \frac{ 2+A^2}{1+A^2} \right) + \tau +1 - \tau \eta^2 \sqrt{ \frac{ 1+A^2}{1+ \eta^2 A^2} } = 0.
			 			\end{equation*}

			 		\subsection*{Phase II}
			 			
			 			From the result above as well as from the analysis of the unitary matrix model, we find a phase transition to a two-cut solution, with a gap opening at $x=0$. The new phase is the conformal image of Phase II of the unitary matrix model.\par
			 			We look for a new eigenvalue density, supported on $[-A,-B] \cup [B,A]$. The result is 
			 			\begin{align}
			 				 \rho^{\mathrm{p.}} _{\mathrm{II}} (x) & =  2 \sqrt{ (A^2-x^2)(x^2-B^2) } ~ \vert x \vert  ~   \left[   - 2 Y  \frac{ x^2 (A^2 +B^2 +2) +3 (A^2 + B^2) +2 A^2 B^2 +4 }{\left[ (1+A^2)(1+B^2)  \right]^{\frac{3}{2}} (1+x^2)^2 }   \right. \notag \\
			 				 	&  \left.   - \frac{\tau+1}{\sqrt{ (1+A^2)(1+B^2)} (1+x^2)}  + \frac{  \tau \eta^4 }{ \sqrt{ (1+ \eta^2 A^2)(1+ \eta^2 B^2)} (1+ \eta^2 x^2)} \right] .
			 			\end{align}
			 			The parameters $A$ and $B$ are fixed by normalization,
			 			\begin{equation*}
			 				- Y \frac{ (A^2-B^2)^2}{ [(1+A^2)(1+B^2) ]^{\frac{3}{2}} } - (\tau+1) \left(  \frac{A^2 + B^2 + 2 }{ \sqrt{(1+A^2)(1+B^2)} } -1 \right) + \tau \left(  \frac{\eta^2 A^2 + \eta^2 B^2 + 2 }{ \sqrt{(1+\eta^2 A^2)(1+\eta^2 B^2)} } -1 \right) = 1 ,
			 			\end{equation*}
			 			and by an additional self-consistency condition on $\omega (z)$,
			 			\begin{equation*}
			 				2 Y \frac{ A^2 +B^2 +2}{(1+A^2)(1+B^2)} + \tau +1 - \tau \eta^2 \sqrt{ \frac{ (1+ A^2) (1+B^2)}{ (1+ \eta^2 A^2)(1+ \eta^2 B^2) } } =0 ,
			 			\end{equation*}
			 			which reproduces the criticality condition for $B \to 0$.\par
			 			\medskip
			 			This is not the end of the story for Phase II. Indeed, fluctuations in the number of eigenvalues in each cut may contribute at leading order in the evaluation of observables \cite{Bonnet:2000dz}. However, for the symmetric two-cut solution we find out that this is not the case, as proved in Appendix \ref{app:fillingfrac}.

			 		\section{Wilson loops and instantons}
			 		\label{sec:WLinst}
			 		
			 		We continue the investigation of the features of the phase transitions and establish their order by evaluating the vacuum expectation value (vev) of the Wilson loop in the fundamental representation. Moreover, we further discuss the different physics of the various transitions by looking at the different contributions by instantons.\par
			 		
			 	\subsection{Wilson loops}
			 		Wilson loops are order operators in gauge theories that, for simple connected gauge group, describe the holonomy of the gauge connection around a closed path. For our one-plaquette model, we consider the Wilson loop in the fundamental representation wrapping the plaquette, and compute its vev. It is given by 
			 			\begin{equation*}
			 				\langle \mathcal{W} \rangle = \left\langle \frac{1}{2N} \tr U  +  \frac{1}{2N} \tr U^{\dagger} \right\rangle = \left\langle \frac{1}{N} \sum_{j=1} ^{N} \cos \theta_j \right\rangle 
			 			\end{equation*}
			 			with the average taken in the unitary ensemble \eqref{MM0}. We use the eigenvalue density at large $N$ found in each phase to evaluate the Wilson loop.\par

					\subsection*{Wilson loops: Generalities}
						From the matrix model \eqref{MM0} we immediately get the relation 
						\begin{equation*}
							\langle \mathcal{W} \rangle = \frac{1}{2N} \frac{1}{ \mz} \frac{ \partial \ }{ \partial (N Y) } \mz = \frac{1}{2} \frac{ \partial \mf }{ \partial Y }. 
						\end{equation*}
						Therefore, all the information about the order of the transition can be extracted from the Wilson loop vev. This is precisely what we expect from an order parameter, and follows from the Wilson loop belonging to the class of order operators of QCD$_2$.\par
						Being $\rho (\theta)$ continuous on the whole $\mathfrak{M}$, the Wilson loop vevs are continuous as well, implying that every phase transition we find must be at least second order.

			 			\subsection*{Wilson loops: Evaluation}
			 			We focus now on the Wilson loop vev at large $N$. In the ungapped phase we find 
			 			\begin{equation}
							\langle \mathcal{W} \rangle_{0} = \int_{- \pi} ^{\pi } \frac{ \dd \theta}{2 \pi } \rho_{0} (\theta) ~ e^{\ii \theta} = Y - \frac{\tau}{\mu} .
						\end{equation}
						This reproduces the GWW result as $\tau \to 0$, but also as $\mu \to \infty$, as expected when the matter becomes non-dynamical. For a Wilson loop winding $k>1$ times around the plaquette, either in clockwise or anti-clockwise direction, we get 
						\begin{equation*}
							\langle \mathcal{W}^{k} \rangle_{0} = - \frac{\tau}{\mu^{k}} .
						\end{equation*}\par
						In Phase Ia the Wilson loop vev is 
						\begin{align}
							\langle \mathcal{W} \rangle_{\mathrm{Ia}} = \int_{- \pi} ^{\pi } \frac{ \dd \theta}{2 \pi } \rho_{\mathrm{Ia}} (\theta) ~ e^{\ii \theta} & = \frac{2}{\pi} \int_{y_0} ^{1} \dd y ~ y \sqrt{ \frac{y-y_0}{1-y} }  \left[ Y - \frac{\tau \mu (\mu-1)}{ \sqrt{1 + \mu^2 - 2 \mu y_0} ( 1+ \mu^2 - 2 \mu y ) } \right] \notag \\
							& =  Y  \frac{(1-y_0)(3+y_0) }{ 4 }  - \frac{\tau}{2 \mu } \left[  \mu ^2 +1 + \frac{  1 + \mu (\mu -1 ) y_0 - \mu^3  }{ \sqrt{1 + \mu^2 - 2 \mu y_0 } }  \right] 
						\end{align}
						where we have used the change of variables $y = \cos \theta$, with $y_0 = \cos \theta_0$. The value of $y_0$ as a function of the gauge theory parameters is known from \eqref{eq:condy0Ia}.\par
						The Wilson loop vev in Phase Ib is obtained likewise, 
						\begin{equation}
							\langle \mathcal{W} \rangle_{\mathrm{Ib}} =  Y  \frac{(1+\tilde{y}_0)(3-\tilde{y}_0) }{ 4 }  - \frac{\tau}{2 \mu } \left[  \mu ^2 +1 + \frac{   \mu (\mu +1 ) \tilde{y}_0 - \mu^3 -1 }{ \sqrt{1 + \mu^2 - 2 \mu \tilde{y}_0 } }  \right] ,
						\end{equation}
						where $\tilde{y}_0 =\cos \tilde{\theta}_0$.\par
						To study the derivative of $\langle \mathcal{W} \rangle_{\mathrm{Ia}}$ and establish the order of the phase transition, it suffices to notice that 
						\begin{equation*}
							\frac{ \dd \ }{\dd Y } \langle \mathcal{W} \rangle_{\mathrm{Ia}} = 1 + \left[  \frac{ Y}{4} \left( -2 y_0 - 2 \right) + \frac{\tau}{2} \cdot \frac{(\mu -1) \mu  (y_0+1)}{\left(1+ \mu ^2-2 \mu  y_0 \right)^{3/2}}  \right]  \frac{ \partial y_0 }{\partial Y } .
						\end{equation*}
						This implies 
						\begin{equation*}
							\lim_{y_0 \to -1} \frac{ \dd \ }{\dd Y } \langle \mathcal{W} \rangle_{\mathrm{Ia}} = 1 
						\end{equation*}
						which matches the derivative of $\langle \mathcal{W} \rangle_0$. The computations are identical for the transition between Phase 0 and Phase Ib. Taking a further derivative, $\frac{ \dd^2  \ }{\dd Y^2 } \langle \mathcal{W} \rangle$ vanishes identically in Phase 0, but does not vanish at the critical loci when computed in Phases Ia and Ib.\par
						We conclude that the Wilson loop vev is an order parameter of class $C^1$ at the critical surfaces $Y = Y_{\mathrm{cr,a}} (\tau, \mu)$ and $Y = Y_{\mathrm{cr,b}} (\tau, \mu)$, thus the system shows a pair of third order phase transitions. In particular, both the GWW transition \cite{Gross:1980he,Wadia:2012fr} and the transition in \cite{Baik} are special points on the critical locus of the present model.\par
						\medskip
						Crossing from a one-cut to a two-cut phase, the first derivative of the Wilson loop is not protected. Indeed, in Phase II the derivative of the Wilson loop vev has the schematic form 
						\begin{equation}
						\label{eq:DWDLambdaPhII}
						    \frac{ \dd \ }{ \dd  Y } \langle \mathcal{W} \rangle_{\mathrm{II}} = \int_{y_0} ^{\tilde{y}_0} y \frac{\partial  \ }{ \partial Y} f(y , y_0, \tilde{y}_0) \dd y + \frac{ \partial y_0}{ \partial Y } \int_{y_0} ^{\tilde{y}_0} y \frac{ \partial \ }{\partial y_0 }  f (y , y_0, \tilde{y}_0) \dd y + \frac{ \partial \tilde{y}_0}{ \partial Y } \int_{y_0} ^{\tilde{y}_0} y \frac{ \partial \ }{\partial \tilde{y}_0 }  f (y , y_0, \tilde{y}_0) \dd y ,
						\end{equation}
						with the first term coming from the derivative of the explicit dependence on $Y$, and the other two from the dependence on $Y$ through $y_0 $ and $\tilde{y}_0$. The integrand evaluated at the endpoint vanishes, hence those contributions do not appear.\par
						In \eqref{eq:DWDLambdaPhII}, $f(y,y_0, \tilde{y_0})$ is known explicitly from Section \ref{sec:LargeNphases}, 
						\begin{equation*}
						    f(y,y_0, \tilde{y_0}) = \frac{2}{\pi} \sqrt{\frac{(y-y_0) (\tilde{y}_0 -y)}{(1+y)(1-y)}} \left[ -Y + \frac{\tau \mu (\mu+1)(\mu -1)}{\sqrt{(1 + \mu^2 - 2 \mu y_0 )(1 + \mu^2 - 2 \mu \tilde{y}_0 )}(1 + \mu^2 - 2 \mu y )} \right]  ,
						\end{equation*}
						but the difficulty comes from the only implicit knowledge of the dependence of $y, \tilde{y}_0$ on $Y$.\par
						Passing from Phase II to Phase Ia, the first term in \eqref{eq:DWDLambdaPhII} matches continuously with the corresponding expression in Phase Ia, as $\tilde{y}_0 \to 1$. The integral in the second summand in \eqref{eq:DWDLambdaPhII} also agrees with the corresponding contribution in Phase Ia at $\tilde{y}_0 \to 1$. Both facts follow from 
						\begin{equation*}
						    \lim_{\tilde{y}_0 \to 1} y_0 \vert_{\mathrm{II}} =  y_0 \vert_{\mathrm{Ia}} .
						\end{equation*}
						Moreover, the symmetries of the integrand allow to combine the third term in \eqref{eq:DWDLambdaPhII} with the second term, in a simpler expression. Moreover, the symmetric form of the equations fixing $y_0, \tilde{y}_0$ can be used to show that 
						\begin{equation*}
						    \frac{\partial \tilde{y}_0}{\partial Y} = \left. \frac{\partial y_0}{\partial Y} \right\rvert_{\tilde{y}_0 \leftrightarrow y_0} .
						\end{equation*}
						By this we mean that the expressions on the two sides agree upon exchanging all $\tilde{y}_0$ with $y_0$.\par
						Due to the complicated dependence on the parameters, the derivatives of the boundaries $y_0, \tilde{y}_0$ are not continuous at the transition point. The differentiability of $ \langle \mathcal{W} \rangle$ above followed by the vanishing of the term multiplying such derivatives. This does not happen for the transition from a two-cut to a one-cut phase. Therefore, the sum of the second and third terms in \eqref{eq:DWDLambdaPhII} gives an obstruction to the differentiability of $ \langle \mathcal{W} \rangle$, so we expect a second order transition. In a sense, the obstruction arises from taking a limit that breaks explicitly the $y_0 \leftrightarrow \tilde{y}_0$ symmetry of Phase II.\par
						The proof is very similar for the transition from Phase II to Phase Ib or from Phase III to either Phase Ia or Ib.\par
						The argument fails at the critical surface $Y_{\mathrm{cr,c}}$ and at the multi-critical point at which $Y_{\mathrm{cr,a}} = Y_{\mathrm{cr,b}}$. Indeed, when passing directly from Phase 0 to a two-cut phase, the simplifications that arise from closing both gaps simultaneously imply that the Wilson loop vev is $C^1$. This is consistent with the observation of the previous paragraph, as these transitions preserve the $\mathbb{Z}_2$-symmetry of the two-cut phase.\par

						\subsection{Phase structure and remarks} 
						Summing up the results extracted from the analysis of Wilson loop vev, we find that 
						\begin{itemize}
							\item the transition from Phase 0 to any other phase is third order, but 
							\item the transition from a one-cut to a two-cut phase is second order.
						\end{itemize}
						In the rest of this subsection we gather comments on various aspects of the phase structure we uncovered, insisting on the role of the second order phase transitions.
						\begin{rmk}
						    As obtained in the previous subsection, the second order discontinuities are finite jumps, not divergences. The correlation lengths remain finite at each transition. These finite discontinuities vanish in the limit $\vert \mu \vert \to 1$.
						\end{rmk}
						
						\subsection*{Metastability}
						While the third order transitions we find are a continuation of the GWW transition in $\mathfrak{M}$, it is worth to further comment on the second order transitions we obtain. The phase transition to a two-cut solution happens slightly beyond the values of $\tau$ where the potential develops a double-well structure. The proposal in \cite{Hanada:2019kue} states that a second order transition can be associated with tunneling from a metastable vacuum to a stable one. Our analysis confirms that picture in the one-plaquette model we consider. In Section \ref{sec:instantons} we study instanton effects, expanding this discussion leading to a further refined distinction between second and third order phase transitions in this model, from the instantonic point of view.\par

                        \subsection*{Critical behaviour}
                        It is worthwhile to notice that the phase diagram in Figure \ref{fig:Phasediag} resembles that in \cite{Demeterfi:1990au}, where a unitary matrix model with potential $Y_1 \cos (\theta ) + Y_2 \cos (2 \theta )$ was analyzed.\par
                        The critical behaviour close to a transition to a two-cut phase in our model differs from that found in similar models in the literature, for matrix models with potentials of the form $\sum_{n=1} ^{K} Y_n \cos (n \theta)$. This is so because the potential in \eqref{MM0} includes both a polynomial and a logarithmic part, requiring different scaling approaching the critical regime from the two-cut phase. This distinction, however, fades away approaching the multicritical point.

						\subsection*{Double-scaling limit}
					       The statements above can be refined exploiting the double-scaling limit.\par 
					       In particular, we can zoom in the critical regime, tuning $Y$ towards a transition to the ungapped phase. In the double-scaling limit, the dynamics is governed by Painlev\'{e} II equation. The proof follows from \cite{BDJ,Baik} with minimal variations. An alternative proof can be given using orthogonal polynomials \cite{Periwal:1990gf}. We have checked explicitly that, in the double-scaling limit, the problem reduces to the analogous one for the pure GWW model.\par
					       For the transition from a one-cut to a two-cut phase, however, there is no double scaling that gives Painlev\'{e} II.
					       
					   \subsection*{Other gauge groups}
					        Throughout the work, we focus on gauge theories with gauge group $U(N)$ or $SU(N)$. Nevertheless, by direct computation or by universality arguments, it can be shown that the phase diagram of Figure \ref{fig:Phasediag} and the associated phase transitions carry over to theories with gauge group $G$, that is to say, matrix models like \eqref{MM0} with the integration over $U(N)$ replaced by integration over $G$, for any
					        \begin{equation*}
					            G \in \left\{   SO(2N), SO(2N+1), Sp(N), O^{-} (2N), O^{-} (2N+1) \right\} .
					        \end{equation*}
					        Here, $O^{-} (n) \subset O(n)$ consists of $n \times n$ orthogonal matrices with determinant $-1$, and $Sp(N)$ is the compact symplectic group.

						\subsection{Continuum limit and $\beta$-function}
						\label{sec:contlimit}

						The $\beta$-function of the theory, as a function of the 't Hooft coupling $\lambda$, can be computed using the chain rule through \cite{Gross:1980he}
						\begin{equation}
						\label{eq:defbeta}
							\beta (\lambda ) = 2 \lambda^2 \frac{ \langle \mathcal{W} \rangle \log  \langle \mathcal{W} \rangle }{ \frac{ \partial \ }{\partial Y}  \langle \mathcal{W} \rangle } .
						\end{equation}\par
						This quantity can be used to test whether our model reproduces the expected features of QCD$_2$ in the continuum limit. The fixed points of the RG flow, that capture the continuum physics, are given by $\beta (\lambda) =0$, which, from \eqref{eq:defbeta}, can only happen at $\langle \mathcal{W} \rangle =0$ or at $\langle \mathcal{W} \rangle =1$.\par
						Direct computations in Phases 0, Ia, Ib, show that only the solution to $\langle \mathcal{W} \rangle =0$ is consistent, while the solution to $\langle \mathcal{W} \rangle =1$ always falls out of the phase in which it has been computed, and thus should be discarded. It is a nice consistency check that the solution to be discarded is precisely the one that would violate Elitzur's theorem \cite{Elitzur:1975im}, and the one to be retained is in agreement with the confining nature of QCD$_2$ \cite{Gross:1980he}.\par
						The continuum limit of a lattice theory consists in sending the lattice spacing to zero while approaching a critical curve \cite{Hernandez:2009zz}. In particular, this requires $\vert \mu \vert \to 1$.\par
						Taking the continuum limit from Phase 0, approaching either $Y_{\mathrm{cr,a}}$ or $Y_{\mathrm{cr,b}}$, we find that the unique consistent solution is $Y=0$ (i.e. $\lambda= \infty$). This is physically meaningful for a toy model of QCD$_2$: the theory flows to a strongly interacting theory in the deep infrared.\par
						Taking the continuum limit from Phase Ia close to the transition to Phase II, we find a trivial solution with $\lambda= 0 = \tau$, describing a theory of free gauge bosons without matter. The continuum limit approaching the critical surface between Phase Ib and Phase II, instead, yields a non-trivial fixed point at 
							\begin{equation*}
    							\lambda = \frac{1}{Y} \approx 121.4 .
	    					\end{equation*}
	    				\begin{rmk}
	    				    The existence of a continuum theory is not established by our analysis, because correlation lengths remain finite. While this has no effect in our model, which consists of a single plaquette, it may (and most likely shall) wash away the fixed point in the continuum limit of a more realistic lattice model.
	    				\end{rmk}

					\subsection{Chiral symmetry breaking}
					\label{sec:chiralSB}
						Let us focus now on the model with fermionic matter. The fermion two-point function is by definition 
						\begin{equation*}
							\langle \bar{\psi}_f \psi_f \rangle = - \frac{ 1}{N \mz} \frac{ \partial \ }{ \partial \mu_f } \mz = -  \frac{ \partial \ }{ \partial \mu_f } \frac{1}{N} \log \mz .
						\end{equation*}
						Due to our degenerate choice of masses, we can only compute the average over flavours of such quantity:
						\begin{equation*}
							\left\langle \frac{1}{K } \sum_{f=1} ^{K} \bar{\psi}_f \psi_f \right\rangle = \frac{1}{\tau} \frac{ \partial \mf }{\partial \mu } .
						\end{equation*}
						In Phase 0 we find 
						\begin{equation*}
							  \frac{1}{\tau} \frac{ \partial \mf_0 }{\partial \mu }  = \frac{2}{\mu^2} \left(  Y  + \frac{\tau \mu}{\mu^2 -1} \right) .
						\end{equation*}
						This quantity diverges as $\mu \to 1$, therefore we expect the chiral symmetry to be spontaneously broken in the continuum, consistently with the analysis of the $\beta$-function in Phase 0.\par
						In Phases Ia and Ib, we can move along $Y =0$ and study the behaviour of $\langle  \frac{1}{K} \sum_f \bar{\psi}_f \psi_f \rangle$ on that subspace of $\mathfrak{M}$. The result is read off directly from \cite{Santilli:2020ueh}: 
						\begin{equation*}
							\left.  \frac{1}{\tau} \frac{ \partial \mf_{\mathrm{Ia}} }{\partial \mu }  \right\rvert_{Y=0} = - \frac{ \mu +1 + 4 \tau }{ 2 \tau \mu (\mu -1)} 
						\end{equation*}
						in Phase Ia, which is non-vanishing and continuous at the transition point.\par
						In Phase Ib we get 
						\begin{equation*}
							\left.  \frac{1}{\tau} \frac{ \partial \mf_{\mathrm{Ib}} }{\partial \mu }  \right\rvert_{Y=0} = - \frac{ 1- \mu + 4 \tau }{ 2 \tau \mu (\mu +1)}  ,
						\end{equation*}						
						again non-vanishing and continuous at the transition point, and goes to $1$ in the $\mu \to 1^{+}$ limit. This latter result, in turn, hints at a transition to a free theory: the free energy of a theory of $K$ free flavours of mass $m$ goes as $\mf_{\text{free}} \propto K m$, whence $\langle \frac{1}{K} \sum_f \bar{\psi}_f \psi_f \rangle \vert_{\text{free}} =1$. Note that this computation has been done at infinite gauge 't Hooft coupling, which has the physical meaning of governing the theory in the deep infrared.\par
						To sum up, we have observed that the phase transition from Phase Ib to Phase 0 is accompanied with spontaneous chiral symmetry breaking.\par

				\subsection{Instantons}
				\label{sec:instantons}
				We discuss non-perturbative effects in the unitary matrix model, coming from unstable saddle point configurations \cite{Marino:2008ya}.\par
				
				An instanton configuration is characterized by a collection of integers $\left\{  N_0, N_1, \dots \right\}$ with $\sum_k N_k =N$. For example, the $d$-instanton configuration is associated with the symmetry breaking pattern 
				\begin{equation*}
					U(N) \rightarrow U(N_0) \times U(N_1) \times \cdots \times U(N_d) ,
				\end{equation*}
				with the eigenvalues $z_j \in \ct$ of $U \in U(N)$ grouped in $d$ different sets, sitting at $d$ different extrema of the potential. For the one-instanton configuration, 
				\begin{equation*}
					\mz_{U(N)} ( \nu ) = \sum_{\ell =0} ^{N} e^{- \nu \ell} \mz_{\ell} 
				\end{equation*}
				with $\mz_{\ell}$ the partition function of a $U(N- \ell) \times U(\ell)$ model, and we have turned on a chemical potential $\nu >0$ for the instanton number.\par
				The $\ell$-sector leads to non-perturbative corrections to the free energy of the matrix model, of the form 
				\begin{equation*}
					e^{-N \ell S_{\text{inst}} ( \vec{\lambda} )}  f_{\ell} ( \vec{\lambda} ) 
				\end{equation*}
				where $\vec{\lambda}$ generically denotes the couplings of the theory, and the functions $\left\{  f_{\ell} \right\}_{\ell}$ admit themselves a $\frac{1}{N}$ expansion.\par
				
			\subsection*{Instanton effects and third order transitions}
				
				Let us consider our model \eqref{MM0} at large $N$ and focus on the one-cut phase, in which the interpretation of instanton effects is more transparent. We discuss them in Phase Ia, being the corresponding analysis in Phase Ib completely analogous. Most of the details are just an extension of the thorough analysis in \cite{Marino:2008ya}.\par
				The contribution of an instanton excitation, obtained moving one eigenvalue from the minimum of $V_{\mathrm{eff}}$ to the maximum at $\theta= \pm \pi $ is found to be 
				\begin{equation*}
				\begin{aligned}
					\pi S_{\text{inst}} & = 2 Y \left[  \sqrt{1 - x_0 ^2} - x_0 ^2 \cosh^{-1} \left( \frac{1}{x_0} \right) \right] \\
					& + \tau \left[  \tanh^{-1} \left( (\mu -1) \sqrt{ \frac{1-x_0 ^2 }{  (\mu-1)^2 + 4 \mu x_0^2 } } \right)  + \frac{\mu -1}{ \sqrt{ (\mu-1)^2 + 4 \mu x_0^2} } \log \left( \frac{x_0}{ 1+\sqrt{1-x_0^2} } \right)\right] 
				\end{aligned}
				\end{equation*}
				where $\cosh ^{-1}$ and $\tanh^{-1}$ are the inverse of the hyperbolic functions, and $x_0 = \sin \frac{\theta_0}{2}$. One of the results in \cite{Marino:2008ya} (already conjectured in \cite{Neuberger:1980qh}) is that the GWW transition is triggered by instantons. We see that the result carries over to the present model, as 
				\begin{equation*}
					\lim_{x_0 \to 1 } S_{\text{inst}} =0 
				\end{equation*}
				and the instanton excitations cease to be suppressed at the critical point when the gap closes.\par
				Analogous conclusions are found if we go to Phase III, in which the effective potential has developed a double well, and consider the instanton configuration with a few eigenvalues taken to the local maximum at $\theta_{\ast}$. Approximating close to the transition to Phase 0, we find 
				\begin{equation*}
					\pi S_{\text{inst}} = \left(  \delta \theta \right)^2 \sin \theta_{\ast} \left[  - Y + \frac{ \tau \mu (\mu +1)}{( \mu -1) \left( 1 + \mu^2 - 2 \mu \cos \theta_{\ast} \right)}   \right] + \mathcal{O} \left( \left(  \delta \theta \right)^3 \right) .
				\end{equation*}
				At the critical surface, $\delta \theta \to 0$ and we find again that the third order transition is triggered by instantons.

				\subsection*{Instanton effects and second order transitions}
\begin{figure}[htb]
			 			\centering
			 				\includegraphics[width=0.5\textwidth]{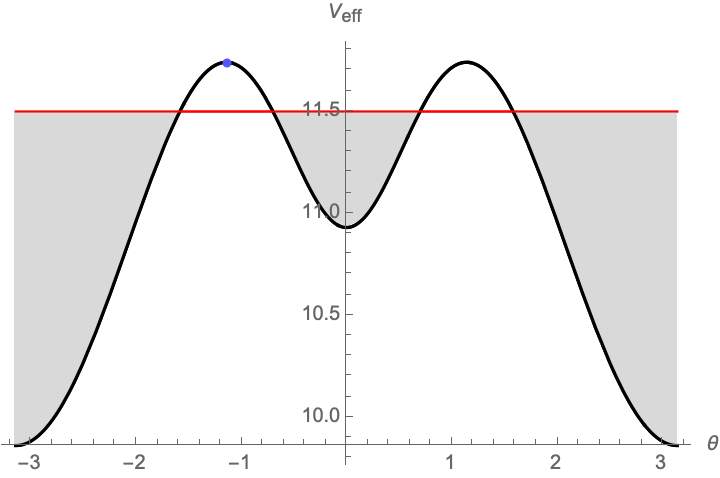}
			 			\caption{Instantons in Phase III. A single eigenvalue (blue dot) is moved on top of the maximum of $V_{\mathrm{eff}}$, while all the others (gray sea) fill the minima.}
			 			\label{fig:instantonIII}
			 			\end{figure}

					We now turn our attention to the analysis of instanton effects in the two-cut phase, starting from Phase III. We consider a single eigenvalue placed on the maximum of the potential, as sketched in Figure \ref{fig:instantonIII}.\par
					
					\par
			 			We find that the instanton action is the sum of two pieces, 
			 			\begin{align*}
			 				S_{\mathrm{inst},L} & = \int_{y_L} ^{y_{\ast}} \frac{ \dd y }{\pi } \sqrt{ \frac{(y -y_L)(y_R-y)}{ 1-y^2 } } \left[  - Y + \frac{ \tau \mu (\mu^2 -1)}{ \sqrt{ (1 + \mu^2 - 2 \mu y_L)  (1 + \mu^2 - 2 \mu y_R)} [ 1 + \mu^2 - 2 \mu y ] }    \right] , \\
			 				S_{\mathrm{inst},R} & = \int_{y_{\ast}} ^{y_R} \frac{ \dd y }{\pi } \sqrt{ \frac{(y -y_L)(y_R-y)}{ 1-y^2 } } \left[  - Y + \frac{ \tau \mu (\mu^2 -1)}{ \sqrt{ (1 + \mu^2 - 2 \mu y_L)  (1 + \mu^2 - 2 \mu y_R)} [ 1 + \mu^2 - 2 \mu y ] }    \right] ,
			 			\end{align*}
			 			where $y_{\ast}= \cos \theta_{\ast}$ and $y_{L,R} = \cos (\theta_{\ast} \pm \delta \theta)$. The two are associated with the eigenvalue escaping from the left and right cut, respectively. There exists a third relevant quantity, namely the tunneling from one cut to the other, 
			 			\begin{equation*}
			 				S_{\mathrm{tunnel}} = \int_{y_L} ^{y_R} \frac{ \dd y }{\pi } \sqrt{ \frac{(y -y_L)(y_R-y)}{ 1-y^2 } } \left[  - Y + \frac{ \tau \mu (\mu^2 -1)}{ \sqrt{ (1 + \mu^2 - 2 \mu y_L)  (1 + \mu^2 - 2 \mu y_R)} [ 1 + \mu^2 - 2 \mu y ] }    \right] .
			 			\end{equation*}
			 			All the three effects are non-perturbatively suppressed by a factor $e^{-N S_{\mathrm{inst}}}$, with $S_{\mathrm{inst}}$ the corresponding action. The three contributions are still suppressed at the critical loci, although the tunneling term will coalesce with one of the other two.\par
\begin{figure}[htb]
			 			\centering
			 				\includegraphics[width=0.5\textwidth]{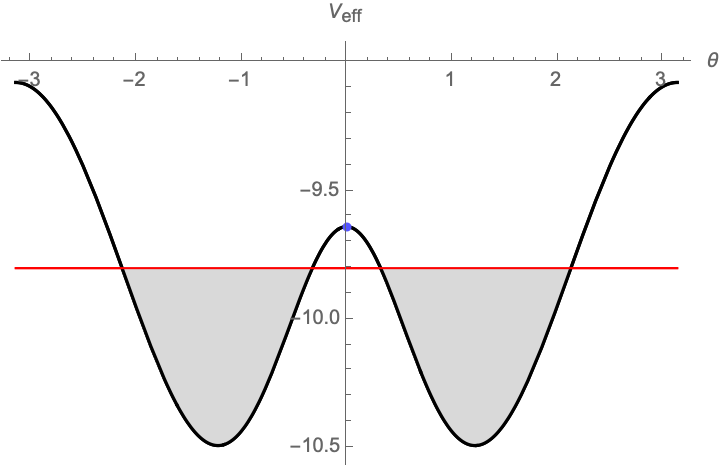}
			 			\caption{Instantons in Phase II. A single eigenvalue (blue dot) is moved on top of the local maximum of $V_{\mathrm{eff}}$ at $\theta=0$, while all the others (gray sea) fill the minima.}
			 			\label{fig:instantonII}
			 			\end{figure}

			 			\medskip
			 			The situation is slightly different in Phase II, where the two wells have equal depth, see Figure \ref{fig:instantonII}.
			 		In this case, $S_{\mathrm{tunnel}}$ is simply twice $S_{\mathrm{inst}}$, and both go to zero as a gap closes. The phase transition takes place when the tunneling between the two cuts ceases to be suppressed in one direction (e.g. passing through $\theta=0$ in Figure \ref{fig:instantonII}) but remains non-perturbative in the other direction (e.g. passing through $\theta=\pi$ in Figure \ref{fig:instantonII}).\par
			 			The picture we infer is that the third order phase transitions are associated with releasing non-perturbative instabilities, while the second order transitions correspond to release only those instabilities in one direction. This is also in agreement with the proposal in \cite{Hanada:2019kue,Hanada:2019rzv} relating second order phase transitions in GWW-type models to partial deconfinement.

				\section{Meromorphic deformation of unitary matrix models}
				\label{sec:merodef}
				
					We now depart from the model \eqref{MM0} with the aim of setting the stage for the study of meromorphic deformations of unitary matrix models, in which the integration contour is deformed in $\C^{\ast}$ and not bound to be the unit circle. This consists of an adaptation of the theory of holomorphic matrix models \cite{Lazaroiu:2003vh} to unitary matrix models and, as we shall show, is instrumental in understanding their phase diagram from new angles. This section can be read independently of the rest of the paper.\par
					A unitary matrix model is characterized by a weight function $e^{-\frac{N}{\lambda} V (z)}$, with $V(z)$ admitting the expansion 
					\begin{equation}
					\label{eq:meroV}
						V (z) = \sum_{n \ge 1 } \left(  \frac{t_n}{n} z^{n} + \frac{t_{-n} }{n } z^{-n} \right) .
					\end{equation}
					The function $e^{- \frac{N}{\lambda} V (z)}$ is singular at $z \in \left\{ 0, \infty \right\} \subset \P1 $ and possibly has other zeros and poles in $\C^{\ast} \cong \P1 \setminus \left\{  0, \infty \right\}$. The Vandermonde determinant appearing in a unitary matrix model is conveniently rewritten in meromorphic form 
					\begin{equation*}
						\prod_{1 \le j < k \le N} (z_j - z_k) \left(\frac{1}{z_j} - \frac{1}{z_k} \right)  =  \left( \prod_{j=1} ^{N} \frac{1}{z_j ^{N-1}}  \right) ~ \prod_{1 \le j < k \le N} (z_j - z_k)^2 .
					\end{equation*}\par
					To deform a unitary matrix model, the integration cycle $\ct ^{N}$ is replaced by any half-dimensional cycle $\mC_N$ in $(\C^{\ast})^{N}$. 
					\begin{defin} Let $N \in \mathbb{N}$, $\lambda \in \C^{\ast}$ and $V(z)$ as in \eqref{eq:meroV}. A meromorphic matrix model $\mz$ is the integral 
					\begin{equation}
					\label{eq:Zmero}
						\mz = \oint_{\mathcal{C}_N} \prod_{1 \le j < k \le N} (z_j - z_k)^2 \prod_{j=1} ^{N}  e^{- \frac{N}{\lambda} V (z_j) } ~ \frac{ \dd z_j}{ 2 \pi \ii  z_j ^{N}}  ,
					\end{equation}
					with integration contour 
					\begin{equation}
					\label{eq:cycle}
						\mC_N = \sum_{\ell} N_{\ell} \mC_{\ell} , \qquad \sum_{\ell} N_{\ell} = N  , \qquad \left[ \mC_{\ell} \right] = \left[ \ct \right] \in H_1 ( \C^{\ast} ) .
					\end{equation}
					\end{defin}
					Condition \eqref{eq:cycle} means that each $\mathcal{C}_{\ell}$ is homotopic to the unit circle in the holed plane $\C^{\ast}$. Dropping it, we may stretch $\mathcal{C}_{\ell}$ along any direction along which, asymptotically, $\Re \frac{1}{\lambda} V (z) >0$. Eventually the one-cycle $\mathcal{C}_{\ell}$ pinches at $z= \infty \in \P1$. To get an honest deformation of a unitary matrix model we do not allow this situation, otherwise we would fall back in the holomorphic deformation of a Hermitian matrix model.\par
					The couplings $\left\{  t_n \right\}$ in \eqref{eq:meroV} are usually subject to reality conditions, such as 
					\begin{equation*}
						t_{-n} = \bar{t}_{n}  , \qquad \forall n \ge 1 ,
					\end{equation*}
					and possibly other relations. We write the constraints collectively as $ \vec{\Phi} \left( \left\{  t_n \right\} \right) =0$. Besides, a rescaling of all $\left\{  t_n \right\}$ together can be reabsorbed in a redefinition of $\lambda$, hence the couplings $\left\{ t_n \right\}$ are homogeneous coordinates on a projective space.\par
					The matrix model \eqref{eq:Zmero} sets a natural stage to complexify the couplings. We denote by $\cT$ the physical space of couplings, namely the collection of independent $\left\{  t_n \right\}$ after imposing the constraints and modulo scaling. More formally, 
					\begin{equation}
					\label{eq:Tspace}
						\cT =   \left(   \left\{   t_n \in \C , \  \forall n \ne 0 \right\} / \C^{\ast} \right)  \cap \left\{  \vec{\Phi} \left( \left\{  t_n \right\} \right) =0  \right\} ,
					\end{equation}
					with the $\C^{\ast}$-action being multiplication of all couplings by a non-vanishing constant. Whenever the constraints $\vec{\Phi}$ can be rewritten in homogeneous form, $\cT$ is a projective variety.\par

						\subsection{Large $N$ limit}
						We are interested in the large $N$ limit of \eqref{eq:Zmero}. Define the effective potential 
						\begin{equation*}
							W (z) := \frac{1}{\lambda } V (z) + \log z .
						\end{equation*}
						At large $N$, the eigenvalues will be gathered around the saddle points of $W (z)$ in $\C^{\ast}$, 
					\begin{equation*}
						W^{\prime} ( z_{\mathrm{sp}, \ell} ) = 0 , \quad \ell = 1, 2, \dots, g +1 .
					\end{equation*}
					Here we are assuming there is a finite number $g +1$ of saddle points $z_{\mathrm{sp}}$. The integration contour $\mC_N$ in \eqref{eq:cycle} can be chosen in such a way that each $\mC_{\ell}$ passes through $z_{\mathrm{sp}, \ell }$. The integers $N_{\ell}$ in \eqref{eq:cycle} then count the number of eigenvalues around the $\ell^{\mathrm{th}}$ saddle point $z_{\mathrm{sp}, \ell }$. At large $N$, the density of eigenvalues will vanish on $\mC_{\ell}$ away from a compact interval $\Gamma_{\ell} \subset \mC_{\ell}$, called a cut, with $z_{\mathrm{sp}, \ell} \in \Gamma_{\ell}$. Therefore  
					\begin{equation*}
						\mathrm{supp} \rho  = \bigcup_{\ell =1} ^{g+1} \Gamma_{\ell} =: \Gamma ,
					\end{equation*}
					and $\rho (z)$ is normalized.
					
					\begin{rmk}\label{rem:1}The requirement that the integration cycle passes through all the $g+1$ saddle points does not fix it uniquely. The shape of each $\mathcal{C}_{\ell}$, and thus of the cuts $\Gamma_{\ell}$ at large $N$, can be homotopically deformed in an open neighbourhood of $z_{\mathrm{sp}, \ell}$, meaning that the matrix model \eqref{eq:Zmero} depends on (up to) $g$ additional parameters.
					\end{rmk}\par
					In the large $N$ limit, the eigenvalue density solves the saddle point equation 
					\begin{equation}
					\label{eq:SPEmero}
						\int_{\Gamma} \frac{   \dd w  }{  2 \pi  }\frac{ \rho (w)}{ z - w } = \frac{1}{2} W^{\prime} (z) ,
					\end{equation}
					where $^{\prime}$ means holomorphic derivative $\frac{\partial \ }{\partial z}$. Here, $\dd w$ is a holomorphic differential on $\Gamma$ and, given any parametrization $w: s \mapsto w (s) \in \Gamma$, $\dd w = \dot{w} (s) \dd s $ is understood, with $\dd s $ the line element and $\dot{w}$ the derivative of the map $w: s \mapsto w (s)$.\par
					Recall the definition of the trace of the resolvent $\omega (z)$ at large $N$:
					\begin{equation*}
						\omega (z) := \int   \dd w  \frac{ \rho (w)}{ z - w }  , \qquad z \in \P1 .
					\end{equation*}
					Equation \eqref{eq:SPEmero} implies that $\omega (z)$ solves \cite{DiFrancesco:1993cyw}
					\begin{equation}
					\label{eq:Riccati}
						\omega (z) ^2 - W^{\prime} (z) \omega (z)  + f (z)  = 0 ,
					\end{equation}
					where 
					\begin{equation*}
						f (z) = \int_{\Gamma} \frac{ W^{\prime} (z) - W^{\prime} (w) }{ z-w } \rho (w) ~ \frac{\dd w }{  2 \pi } .
					\end{equation*}
					Defining 
					\begin{equation}
					\label{eq:defy}
						y (z) = \omega (z) - \frac{1}{2} W^{\prime} (z) ,
					\end{equation}
					\eqref{eq:Riccati} becomes 
					\begin{equation}
					\label{spectral}
						y^2 - \left(  \frac{1}{2 } W^{\prime } (z) \right)^2  + f(z) =  0 .
					\end{equation}
					This equation goes under the name of spectral curve. The steps from \eqref{eq:SPEmero} to \eqref{spectral} are standard and have been applied to holomorphic matrix models since their early days \cite{Dijkgraaf:2002fc} to establish a bridge between matrix models and geometric problems. The novel aspect of \eqref{spectral} compared to the literature is hidden in the form of $W^{\prime} $ and $f$, which in the present case are not ordinary polynomials but Laurent polynomials, or meromorphic functions.\par
					\medskip
					We assume for now that $V(z)$ is a Laurent polynomial on $\P1$, with singularities at $\left\{ 0, \infty \right\}$. The extension to a meromorphic weight function on $\C^{\ast}$ is worked out below. Write 
					\begin{equation}
					\label{defLaurentV}
						V (z) = \sum_{n= - d_{-} +1} ^{d_+ +1} \frac{t_n}{ \vert n \vert } z^{n} \quad \Longrightarrow \quad W ^{\prime} (z) = \frac{1}{\lambda } \sum_{n = - d_{-}} ^{d_{+}} \mathrm{sgn} (n) t_{n+1} z^{n} + \frac{1}{z}  ,
					\end{equation}
					where we assume $d_{-} \ge 2$ (otherwise we get back the known setting).\par
					We need to introduce some notation. Define 
					\begin{equation}
					\label{genus}
						g = d_+ + d_{-} - 1 ,
					\end{equation}
					which agrees with the counting of saddle points above, and also $t_0 = - \lambda $ for later convenience. Besides, denote $\rho_k$ the moments of the eigenvalue density, 
					\begin{equation*}
						\rho_k = \int_{\Gamma}  w^{k}  \rho (w) \dd w , \qquad k \in \mathbb{Z} .
					\end{equation*}
					After some rewriting we get 
					\begin{align}
						W^{\prime} (z) & = \frac{1}{ \lambda z ^{d_{-}}} \sum_{n=0} ^{g  +1} z^{n}t_{n+1-d_{-}} ~\mathrm{sgn} (n-d_{-})  ,  \label{eq:WLaurent}  \\
						f (z) & = \frac{1}{\lambda  z^{d_{-}}} \left[  \sum_{n=0}^{d_{-} -1} z^{n} \left( \sum_{k=-1} ^{n-1} t_{k+2-d_{-}} \rho_{k-n}  \right) + \sum_{n=d_{-} } ^{g} z^{n} \left( \sum_{k=n} ^{g} t_{k+2-d_{-}} \rho_{k-n}  \right)  \right] ,  \label{eq:fLaurent} 
					\end{align}
					where $\mathrm{sgn} (0) =+1$ by convention. The spectral curve takes the schematic form 
					\begin{equation}
					\label{eq:frachyperelliptic}
						y^2 = \frac{ P (z)  }{4 \lambda^2 z^{2 d_{-}}} 
					\end{equation}
					where $P(z)$ is a polynomial in $z$ of degree $\mathrm{deg} (P) = 2 g +2 $, with coefficients read off from \eqref{eq:WLaurent}-\eqref{eq:fLaurent} and that depend on the parameters $\left\{ t_n \right\}$, on $\lambda$ and on the moments $ \left\{ \rho_k  \right\}$.\par
					A major difference with respect to the standard unitary matrix models is that $\left\{  \rho_k \right\}$ are free complex parameters of the theory: they can be tuned deforming $\Gamma$, as discussed in Remark \ref{rem:1}. Recalling that both $z, \lambda \in \C^{\ast} $, it is possible to recast \eqref{eq:frachyperelliptic} in a more standard form $\hat{y} = P(z)$, describing an hyperelliptic complex curve of genus $g$ \cite{Dijkgraaf:2002fc}.\par
					$\Gamma$ is the union of $g+1$ branch cuts stretched between pairs of roots of $P(z)$. The roots of $P(z)$ move inside $\C^{\ast}$ as the parameters are varied.\footnote{Without loss of generality, $ \left\{  0, \infty \right\} \subset \P1$ are not roots of $P(z)$, because they would correspond to a ``non-minimal'' choice of $d_{\pm}$ in \eqref{defLaurentV}. They can be avoided simply defining $\hat{y} = 2 y \lambda  z^{ d_{-} \pm  m} $ with $m $ the multiplicity of the root, and minus (resp. plus) sign if the root is $z=0$ (resp. $z= \infty$).} The coalescence of two roots produces a singularity of the curve \eqref{eq:frachyperelliptic} and corresponds, on the matrix model side, to a phase transitions from a $(g+1)$-cut to a $g$-cut phase, with either
					\begin{itemize}
					    \item two cuts joining, or
					    \item one cut collapsing.
					\end{itemize}\par
					\medskip
					The hyperelliptic curve \eqref{eq:frachyperelliptic} is fibered over the moduli space $\cM$ of the model \eqref{eq:Zmero}, defined as 
					\begin{equation*}
						\cM = \C^{\ast} \times \cT \times \C^{g} ,
					\end{equation*}
					with $\C^{\ast}$ parametrized by $\lambda $, $\cT$ defined in \eqref{eq:Tspace}, and the last factor parametrized by the moments $ \left\{ \rho_k  \right\}$. Note that one of the moments is fixed comparing \eqref{eq:Riccati} with the definition of $\omega (z)$ at $\vert z \vert \to \infty$.\par
					\begin{defin}
					A critical locus $\cC $ is an irreducible component of the locus in $ \cM $ at which two roots of $P(z)$ coalesce.
					\end{defin}
					The critical loci $\cC \subset \mathscr{M}$ necessarily have positive complex codimension, and the hyperelliptic fibration is singular along them. Singularities in higher codimension, placed at the (self-)intersection of critical loci, correspond to multicritical points of the matrix model.\par
					The theory of Abelian differentials provides a suitable framework to analyze the genus $g$ hyperelliptic curve \eqref{eq:frachyperelliptic} \cite{Chekhov:2005kd}. At this stage, the analysis of the spectral curve works exactly as in the holomorphic deformation of Hermitian matrix models, thus we omit the details and refer to \cite{Chekhov:2005kd,Alvarez:2013}.\par
					\begin{rmk}
						We are now in the position to elaborate more on Remark \ref{rem:1}, from a point of view advocated in \cite{Bilal:2005hk}. Let $\left\{ A^{\ell}, B_{\ell} \right\}$ be a basis of one-cycles in the hyperelliptic complex curve. The $A$-cycles are chosen to go around the cuts $\Gamma_{\ell}$. Therefore 
						\begin{equation*}
							\oint_{A^{\ell}} y(z) \dd z = \oint_{A^{\ell}} \omega (z) \dd z = \frac{N_{\ell}}{N} =: \xi_{\ell} .
						\end{equation*}
						The first equality follows from the definition \eqref{eq:defy} noting that $y(z)$ and $\omega (z)$ only differ by a regular term. Introducing chemical potentials for the filling fractions $\xi_{\ell}$ and extremizing the action with respect to these quantities gives their saddle point value as a function on $\cM$. More precisely, one gets a set of equations analytic in the ratios $s_{\ell} := \frac{\xi_{\ell}}{\lambda}$ \cite{Bilal:2005hk}. At this point, it is possible to invert the relations and express the moments $\left\{ \rho_{k} \right\}$ in terms of the complex variables $s_{\ell}$, keeping the latter as free parameters.\par
						Note that only $g$ out of the $g+1$ of both quantities are free.
					\end{rmk}\par
					The study of the phases of the model \eqref{eq:Zmero} leads to a stratification of the parameter space $\mathscr{M}$. We postpone the analysis to Section \ref{sec:stratification}, discussing explicit models first.

                        \subsection*{Genus 0}
						The unique way to obtain a genus 0 spectral curve is from the holomorphic deformation of the CUE. In that case, the model has no couplings and, as opposed to $g \ge 1$, the additional condition derived from the definition of $\omega (z)$ is automatically fulfilled, leaving $\rho_{-1}$ as unique, unconstrained parameter. Then, \eqref{eq:frachyperelliptic} describes a $\mathbb{P}^{1}$ fibered over $\mathbb{C}$. If we try to get a less trivial model by considering the insertion of $\left( \det U \right)^{\tau N }$, the consistency condition, which in $g \ge 1$ fixes one of the $\left\{ \rho_{k} \right\}$, imposes $\tau =0$.

						\subsection{Holomorphic GWW}
						\label{sec:holoGww}
						We now put the machinery at work and revisit the phase structure of the holomorphic GWW model. The phase diagram of this model has been obtained in \cite{Alvarez:2016rmo} for $\lambda \in \R$, while the behaviour at complex coupling has been partially analyzed in \cite{Copetti:2020dil}, although without fully exploiting the holomorphic deformation.\par
						The GWW model has $t_{-1}=t_1 =1$, and $t_{n \ne \pm 1} =0$, whence $d_{+}=0$, $d_{-}=2$, $g=1$ and only the moments $\rho_{-2}, \rho_{-1}$ appear in \eqref{eq:fLaurent}. Fixing $\rho_{-2}$ as a function of $\lambda$ and $\rho_{-1}$, the spectral curve of the holomorphic GWW model is \cite{Alvarez:2016rmo}
						\begin{equation}
						\label{eq:GWWspectral}
							\hat{y}^2 = z^4 + 2 \lambda z^3 + \left[ (\rho_{-1} +1) \lambda^2 -2 \right] z^2 + 2 \lambda z +1 .
						\end{equation}
						It is an elliptic curve. Following the strategy outlined above, we think of \eqref{eq:GWWspectral} as an elliptic fibration over $\C^{\ast} \times \C $, with coordinates on the base $\lambda$ and $\rho_{-1}$, and identify the phase transitions with singularities of the fibration.\par
						The discriminant of \eqref{eq:GWWspectral} is 
						\begin{equation}
						\label{DeltaGWW}
							\Delta = \frac{\lambda^2}{4} \left(  \lambda^2  \rho_{-1}  -4  \right)^2 \left(  \lambda \left(\rho_{-1} +1 \right) -4  \right) \left(  \lambda \left(\rho_{-1} + 1 \right) +4  \right) ,
						\end{equation}
						from which the critical loci are
						\begin{align*}
							\left\{  \Delta =0 \right\}  & = \cC_{1} \cup \cC_{1^{\prime}}  \cup  \cC_{2 } , \\ 
							\cC_{1} & := \left\{ \rho_{-1} = -1 + \frac{4}{\lambda} , \ \lambda \in \C^{\ast}  \right\} , \quad \cC_{1^{\prime}}  := \left\{ \rho_{-1} = -1 - \frac{4}{\lambda} , \ \lambda \in \C^{\ast}  \right\} , \\
							\cC_{2} & := \left\{ \rho_{-1} = \frac{4}{ \lambda^2 } , \ \lambda \in \C^{\ast} \right\} .
						\end{align*}
						In Kodaira's classification \cite{Kodaira}, $\cC_1$ and $\cC_{1^{\prime}} $ are singularities of type I$_1$ and $\cC_2$ is of type I$_2$. The GWW critical points $(\lambda, \rho_{-1})=(\pm 2, 1)$ are singled out as the codimension-two singularities at which $\cC_2$ intersects one of the other two critical curves. Besides, we recognize the elliptic curve \eqref{eq:GWWspectral} as the Seiberg--Witten curve of $\mathcal{N}=2$ supersymmetric four-dimensional $SU(2)$ gauge theory with two flavours \cite{Seiberg:1994aj}.\par
						The original GWW transition is thus, from the perspective of the holomorphic deformation, one of the possible ways to approach the codimension-two singularity from a generic direction. The singularity at the multicritical points $(\lambda, \rho_{-1})=(\pm 2, 1)$ is of Kodaira type III. The corresponding symmetry is $A_1$, which is precisely the symmetry of Painlev\'{e} II, that is known to control the GWW phase transition \cite{Periwal:1990gf,BDJ}. If, instead, we approach the multicritical point not from a generic direction but moving along a critical locus, the singularity type is enhanced to I$_0 ^{\ast}$.\par
						\medskip
						It is possible to allow $t_{-1} \ne t_1$. This corresponds to introduce a $\theta$-term in the GWW lattice action, $\frac{ \theta}{2 \pi} = \frac{t_1 - t_{-1}}{2}$. The procedure goes through with only minor modifications, the unique difference being that the singularities $\cC_1, \cC_{1^{\prime}}$ are placed at $\rho_{-1} = - \frac{1}{t_{-1}} \pm \frac{4}{\lambda \sqrt{t_{-1}}}$, so in particular the $\mathbb{Z}_2$-symmetry is preserved.

						\subsection{Meromorphic deformations}
						\label{sec:meroBaik}
							The formulation can be extended to include weight functions with zeros and poles in $\C^{\ast}$. For concreteness, we consider the illustrative example of our original model \eqref{MM0} at $\lambda^{-1}=0$. In this case 
							\begin{align*}
								W^{\prime} (z) & =  -  \tau \left[ \frac{ 1}{z-\mu } + \frac{1}{z - \mu^{-1}} \right] + \frac{1+ \tau}{z} , \\
								f (z) & = \tau \left[ \frac{ \tilde{\rho}_{-1} (\mu) }{z- \mu} + \frac{ \tilde{\rho}_{-1} (\mu^{-1}) }{z- \mu^{-1}}  \right] - \frac{(1+\tau)}{z} \rho_{-1} .
							\end{align*}
							In the second line, we have defined 
							\begin{equation*}
								\tilde{\rho}_{-1} (\mu) := \int_{\Gamma}  \frac{\rho (w)}{ w - \mu }  \dd w ,
							\end{equation*}
							with, in particular, $\tilde{\rho}_{-1} (0) = \rho_{-1}$. Comparing the definition of $y(z)$ with the spectral curve at large $\vert z \vert$, we find a pair of consistency conditions, fixing $\tilde{\rho}_{-1} $ as a function of the other parameters, 
							\begin{align*}
								\tilde{\rho}_{-1} (\mu) = -\frac{4 \rho_{-1} (\tau +1)+\mu  \tau  (\tau +6)+\mu }{4 \left(\mu ^2-1\right) \tau } .
							\end{align*}
							Note that the two conditions fix $\tilde{\rho}_{-1} (\mu^{\pm 1}) $ independently, and the solutions are consistently mapped into each other under $\mu \leftrightarrow \mu^{-1}$. We get 
							\begin{equation}
								y^2 = \frac{P(z)}{4 z^2 (z-\mu)^2 (z-\mu^{-1})^2} ,
							\end{equation}
							with $P(z)$ a polynomial of degree 4. The spectral curve thus describes again en elliptic fibration over the moduli space $\C^{\ast} \times \left\{ \mu \in \mathbb{P}^1 \ : \ 1 < \vert \mu \vert < \infty \right\} \times \C$, parametrized by $(\tau, \mu, \rho_{-1}) $. The discriminant takes the form $\Delta = \mu^5 (\mu+1)^2 (\mu-1)^2 (\tau+1)^{4} \widetilde{\Delta} $, the last term being a cumbersome polynomial of degree 6 in $\rho_{-1}$, degree 8 in $\tau$ and degree 10 in $\mu$. The critical points of the undeformed model become higher-codimensional singularities, at which two roots of $P(z)$ collide. The collection of all critical loci in this model is 
							\begin{equation*}
								\cC_{4} \cup \cC_{2} \cup \cC_{2^{\prime}} \bigcup_{j=1} ^{6} \cC_{1_{j}} ,
							\end{equation*}
							with the subscript indicating the order of vanishing of $\Delta$ along the component $\cC$. Taking what we have called the continuum limit in Section \ref{sec:contlimit}, that is, sending $\tau \to \tau_{\mathrm{cr}} (\mu)$ and then $\mu \to \pm 1$, with $\rho_{-1}= - \frac{\tau}{\mu}$ set to its undeformed value, yields a non-minimal singularity $\Delta \propto (\mu \pm 1)^{12}$.

						\subsection{Stratification of the moduli space}
						\label{sec:stratification}
						The critical loci and their intersections endow the parameter space $\mathscr{M}$ with additional structure.\par
						The stratification of an algebraic variety $\mathscr{V}$ is a collection of open sets $\left\{ \mathscr{V}_{I} \right\}$, with $\mathscr{V}_0$ a point and $\overline{\mathscr{V}}_{\text{max}} = \mathscr{V}$, with a partial order given by the inclusion of the closures of $\left\{ \mathscr{V}_I\right\}$. The parameter space $\mathscr{M}$ of the model \eqref{eq:Zmero} is the union of 
						\begin{equation*}
						    \mathscr{M}^{\text{reg}} , \ \cC_{I} ^{\text{reg}} , \ \cC_{IJ} ^{\text{reg}} , \ \dots 
						\end{equation*}
						where the superscript means the regular part, and $\cC_{IJ} = \cC_{I} \cap \cC_{J}$, and so on. The inclusion relations $\overline{\cC^{\text{reg}} _{IJ}}  = \overline{\cC_{I}} \cap   \overline{\cC_{J}} \subset \overline{\cC_I} $ are obvious.\par
						The partial order can be represented with the aid of a Hasse diagram:
						\begin{equation*}
						\begin{tikzpicture}
						    \node (mreg) at (0,0) {$\mathscr{M}^{\text{reg}}$};
						    \node (ci) at (-2,-1) {$\mathscr{C} _{I} ^{\text{reg}}$};
						    \node (cj) at (-1,-1) {$\mathscr{C} _{J} ^{\text{reg}}$};
						    \node[] (d1) at (0.5,-1) {$\cdots$};
						    \node (ck) at (2,-1) {$\mathscr{C} _{K} ^{\text{reg}}$};
						    \node (cij) at (-1.5,-2) {$\mathscr{C} _{IJ} ^{\text{reg}}$};
						    \node (cik) at (1,-2) {$\mathscr{C} _{IK} ^{\text{reg}}$};
						     \node[] (d2) at (-0.5,-2) {$\cdots$};
						     \node[] (v1) at (-1.5,-2.5) {$\vdots$};
						     \node[] (v2) at (0,-2.5) {$\vdots$};
						     \node[] (v3) at (1.5,-2.5) {$\vdots$};
						    \draw  (mreg) -- (ci);
						    \draw  (mreg) -- (cj);
						    \draw  (mreg) -- (ck);
						    \draw (ci) -- (cij);
						    \draw (cj) -- (cij);
						    \draw (ci) -- (cik);
						    \draw (ck) -- (0,-2);
						    \draw (ck) -- (cik);
						  
						\end{tikzpicture}
						\end{equation*}
						In general, this does not define a full-fledged stratification of $\mathscr{M}$ because multiple final points may exist. Nonetheless, whenever the potential \eqref{eq:meroV} has a $\mathbb{Z}_2$-symmetry, the Hasse diagram inherits it. This $\mathbb{Z}_2$-symmetry acts as an automorphism of the Hasse diagram, which is mapped into itself under reflection along the vertical axis. By construction, the Hasse diagram resulting from folding the initial diagram via this $\mathbb{Z}_2$-symmetry determines a stratification of $\mathscr{M}/\mathbb{Z}_2$.\par
						We draw the Hasse diagram of the holomorphic GWW model of Section \ref{sec:holoGww}:
						\begin{equation}
						\label{HasseGWW}
						\begin{tikzpicture}
						    \node (mreg) at (0,0) {$\mathscr{M}^{\text{reg}}$};
						    \node (c1) at (-2,-1.5) {$\mathscr{C} _{1} ^{\text{reg}}$};
						    \node (c2) at (0,-1.5) {$\mathscr{C} _{2} ^{\text{reg}}$};
						    \node (c3) at (2,-1.5) {$\mathscr{C} _{1^{\prime}} ^{\text{reg}}$};
						    \node (p1) at (-1,-3) {$ \lambda = 2 $};
						    \node (p2) at (1,-3) {$\lambda = -2$};
						   \draw (mreg) -- (c1);
						   \draw (mreg) -- (c2);
						   \draw (mreg) -- (c3);
						   \draw (c1) -- (p1);
						   \draw (c2) -- (p1);
						   \draw (c2) -- (p2);
						   \draw (c3) -- (p2);
						   \draw[red,thin,dashed] (mreg) -- (0,-3);
						 \end{tikzpicture}
						\end{equation}
						The diagram of the meromorphic model of Section \ref{sec:meroBaik} is schematically
						\begin{equation*}
						\begin{tikzpicture}
						    \node (mreg) at (0,0) {$\mathscr{M}^{\text{reg}}$};
						    \node (c2a) at (-2,-1.5) {$\mathscr{C} _{2} ^{\text{reg}}$};
						    \node (c4) at (0,-1.5) {$\mathscr{C} _{4} ^{\text{reg}}$};
						    \node (c2b) at (2,-1.5) {$\mathscr{C} _{2^{\prime}} ^{\text{reg}}$};
						    \node (c13a) at (-4,-1.5) {$\mathscr{C} _{1_{3}} ^{\text{reg}}$};
						    \node (c13b) at (4,-1.5) {$\mathscr{C} _{1_{4}} ^{\text{reg}}$};
						    \node (c12a) at (-5,-1.5) {$\mathscr{C} _{1_{2}} ^{\text{reg}}$};
						    \node (c12b) at (5,-1.5) {$\mathscr{C} _{1_{5}} ^{\text{reg}}$};
						    \node (c11a) at (-6,-1.5) {$\mathscr{C} _{1_{1}} ^{\text{reg}}$};
						    \node (c11b) at (6,-1.5) {$\mathscr{C} _{1_{6}} ^{\text{reg}}$};
						    \node (p1) at (-1,-3.3) {$ \mu = 1 $};
						    \node (p2) at (1,-3.3) {$\mu = -1$};
						    \node (d1) at (-3,-2) {$\vdots$};
						    \node (d3) at (3,-2) {$\vdots$};
						    \node (e) at (0,-2.5) {};
						     \node (f) at (-2,-2.5) {};
						      \node (g) at (2,-2.5) {};
						   \draw (mreg) -- (c11a);
						   \draw (mreg) -- (c11b);
						    \draw (mreg) -- (c12a);
						   \draw (mreg) -- (c12b);
						    \draw (mreg) -- (c13a);
						   \draw (mreg) -- (c13b);
						   \draw (mreg) -- (c2a);
						   \draw (mreg) -- (c2b);
						   \draw (mreg) -- (c4);
						   \draw (f) -- (p1);
						   \draw (e) -- (p1);
						   \draw (g) -- (p2);
						   \draw (e) -- (p2);
						   \draw[red,thin,dashed] (mreg) -- (0,-3.3);
						 \end{tikzpicture}
						\end{equation*}
						The vertical red, dashed line is there to emphasize the $\mathbb{Z}_2$ reflection symmetry. Folding the diagram along that line yields the stratification of $\mathscr{M}/\mathbb{Z}_2$.\par
						
						\begin{rmk}
						    The results of Section \ref{sec:holoGww} with $t_{-1} \ne t_{1}$ show that, even for models in which a $\mathbb{Z}_2$ reflection symmetry is not manifest from the potential but emerges at large $N$, the $\mathbb{Z}_2$-folding yields a stratified moduli space.
						\end{rmk}
						
						\subsection*{Symplectic singularities}
						Recall that $H_{1} (\Sigma_g, \mathbb{C})$, the first homology group of a hyperelliptic curve $\Sigma_g$ of genus $g$, is a symplectic space. The $A$- and $B$-cycles that we have implicitly used in the study of the spectral curve \eqref{spectral} can be chosen to be Darboux coordinates in $H_{1} (\Sigma_g, \mathbb{C})$. Moving along $\mathscr{M}^{\text{reg}}$ corresponds to vary the symplectic structure without changing the topology of $\Sigma_g$. At the critical loci $\cC_{I}$, however, either 
						\begin{itemize}
						    \item a $B$-cycle collapses, or 
						    \item an $A$-cycle collapses.
						\end{itemize}
						Both situations correspond to a singularity of the symplectic form. Therefore, the analysis of the phase structure of the meromorphic matrix models can be rephrased in terms of symplectic singularities in the sense of Kaledin \cite{Kaledin}.\par
						The appearance of symplectic singularities is not entirely unexpected. The consideration of holomorphic matrix models in their large $N$ limit lead to the Seiberg--Witten curves \cite{Seiberg:1994aj} of certain  $\mathcal{N}=2$ gauge theories \cite{Dijkgraaf:2002fc}. The so-called Coulomb branch of these theories is a symplectic singularity and is stratified \cite{Argyres:2020wmq}. In fact, the use of Hasse diagrams in the present work was inspired by \cite{Bourget:2019aer,Grimminger:2020dmg}.\par
						\medskip
						As a final remark, we emphasize that the structure uncovered in this section is not specific of the meromorphic matrix models. The  parameter spaces of unitary or Hermitian matrix models inherit it, as they can be realized as slices inside the parameter space of our meromorphic models.\par
						As an example, the phase diagram of the GWW model is 
						\begin{equation}
						\label{HasseRealGww}
						\begin{tikzpicture}
						    \node (mreg) at (0,0) {$\bullet$};
						    \node (p1) at (-1,-1.5) {$ \bullet $};
						    \node (p2) at (1,-1.5) {$\bullet $};
						   \draw (mreg) -- (p1);
						   \draw (mreg) -- (p2);
						   \draw[red,thin,dashed] (mreg) -- (0,-1.5);
						   \node[anchor=west] (mgww) at (0.5,0) {$\mathscr{M}^{\mathrm{reg}} \vert_{\mathrm{GWW}} =  \left\{  \lambda \in \mathbb{R} \setminus \left\{0 \right\} , \ \lambda^2 \ne 4\right\}  $};
						   \node[anchor=east] (c1gww) at (-1.2,-1.5) {$\cC_1 \vert_{\mathrm{GWW}} = \left\{ \lambda=2 \right\}$};
						    \node[anchor=west] (c2gww) at (1.2,-1.5) {$\cC_{1^{\prime}} \vert_{\mathrm{GWW}} = \left\{ \lambda=-2 \right\}$};
						 \end{tikzpicture}
						\end{equation}
						It is found by fixing $\rho_{-1} =1$, taking the slice $\lambda \in \R \setminus \left\{0 \right\}$ and identifying the intersection of such subspace with the strata in \eqref{HasseGWW}.\footnote{Accidentally, this is precisely the Hasse diagram of the reduction to three dimensions of the $SU(2)$ theory with two flavours, captured by the holomorphic GWW of Section \ref{sec:holoGww}, cf. \cite[Eq.(4.2)]{Grimminger:2020dmg}. It should be stressed, however, that the strata in \eqref{HasseRealGww} are real, not hyperK\"{a}hler.}

							\section{Outlook}
							\label{sec:outlook}
							
							We conclude by commenting, in a qualitatively and non-exhaustive fashion, on avenues that we have considered at some point but not pursued here.\par
							It would be interesting to know if the results obtained have a meaning from the point of view of integrable systems such as the Schur flow. While both the matrix models considered and the study of such flows have in common an associated system of orthogonal polynomials, the way this association actually works is quite different. For example, the recurrence coefficients of the polynomials, central in the integrable systems description, are not directly relevant in the type of matrix model analysis presented.\par
							On the other hand, for what concerns matrix models on the real line, the spectral properties of Jacobi matrices can be more directly related to matrix models, since it is known that, under rather general conditions, a suitably normalized counting measure of the zeroes of the orthogonal polynomials converges weakly, in the large $N$ limit, to the density of states of the matrix model \cite{PasturShbook}. 
							With this in mind, a question would be whether our results have any implication in the study of the spectral properties of CMV matrices \cite{nenciu2006cmv}, for example. The large $N$ planar limits taken make this possibility not obvious.\par
							Another reason to further study any eventual implications of the planar limit and the ensuing phase structure, from the point of view of integrable systems, would be the connection obtained, presented in Section \ref{sec:stratification}, between the phase diagram of meromorphic matrix models and the symplectic foliation of singular varieties \cite{Kaledin}. Again, the 't Hooft scaling of the couplings involved at large $N$ obscures the relation between the symplectic structures we naturally find and those in the integrability literature \cite{RowanNenciu}.\par
							Also, from a mathematical point of view, it would be interesting if proofs of the order of the phase transition, in particular for the second order phase transitions, can be obtained in alternative or more rigorous ways.\par
							
							Regarding more physical considerations, when discussing the model with fermionic matter and its interpretation in terms of chiral symmetry breaking, it is worth mentioning that recently \cite{Castorina:2020vbh}, the chiral symmetry breaking phase transition in four-dimensional QCD has been studied from the point of view of thermodynamic geometry \cite{Ruppeiner:1979,Ruppeiner:1995}. The argument is based on the observation that the grand canonical partition function 
						\begin{equation*}
							\widehat{\mz} (v) = \sum_{N=0} ^{\infty} e^{-N v } \mz_{U(N)} , \qquad v >0 
						\end{equation*}
						determines a metric $g^{\mathrm{thermo}}$ on a two-dimensional parameter space with coordinates $(\mf, v)$ \cite{Ruppeiner:1979}, where $\mf$ is the free energy and $v$ the grand canonical chemical potential. Then, a second order phase transition is triggered by the instability at $ \det g^{\mathrm{thermo}} =0$.\par
						Any eventual use of this observation or other ideas from information geometry to further understanding phases in matrix models would be of interest.\par

				\vspace{0.25cm}
\subsection*{Acknowledgements}

We thank Jorge Russo for a careful reading and valuable commentaries. The work of LS is supported by the Funda\c{c}\~{a}o para a Ci\^{e}ncia e a Tecnologia (FCT) through the doctoral grant SFRH/BD/129405/2017. The work is also supported by FCT Project PTDC/MAT-PUR/30234/2017.

			\begin{appendix}
			
			\section{Technical details of the solution}

			\subsection{Exact expressions at finite $N$ via Toeplitz determinants}
			\label{app:finiteN}
		    Table \ref{tab:ZexactNK} collects the explicit expressions for $\mathcal{Z}_{U(N)} ^{\epsilon=+1, K}$ for the first few values of $N$ and $K$, computed using the Toeplitz determinant formulation detailed in Section \ref{sec:exact}.

				\begin{table}[thb]
				\centering
				\begin{tabular}{c | c | l }
				 $N$ & $K$ & $\mz_{U(N)} ^{+1,K}$ \\
				 \hline \hline
				 1 & 1 &   $ {\scriptstyle \left(\mu ^2+1\right) I_0(2 Y )-2 \mu  I_1(2 Y ) } $ \\
				 \hline
				 1 & 2 &  $ {\scriptstyle \left(\mu ^4+6 \mu ^2+1\right) I_0(2 Y )-\frac{2 \mu }{Y} \left(2 Y  \left(\mu ^2+1\right)+\mu \right) I_1(2 Y )} $ \\
				 \hline
				 1 & 3 &   $ \begin{aligned}   {\scriptstyle  \frac{1}{Y^2} } & {\scriptstyle  \left[   Y  \left(Y  \left(\mu ^6+15 \mu ^4+15 \mu ^2+1\right)+4 \mu ^3\right) I_0(2 Y )  \right. }  \\   &{\scriptstyle  \left. -2 \mu  \left(Y ^2 \left(3 \mu ^4+10 \mu ^2+3\right)+3 Y  \left(\mu ^3+\mu \right)+2 \mu ^2\right) I_1(2 Y )  \right] }  \end{aligned}  $ \\
				 \hline
				 2 & 1 &   $ \begin{aligned} & {\scriptstyle  2 \left(\mu ^4+\mu ^2+1\right) I_0(4 Y )^2-4 \mu  I_0(4 Y ) \left(\left(\mu^2+1\right) I_1(4 Y )+\mu  I_2(4 Y )\right)   -2 \left(\mu ^2-1\right)^2 I_1(4 Y )^2 } \\ & {\scriptstyle  -2 \mu ^2 I_2(4 Y )^2+4 \mu  \left(\mu ^2+1\right) I_1(4 Y ) I_2(4 Y )   } \end{aligned} $  \\
				 \hline
				 2 & 2 &   $ \begin{aligned} {\scriptstyle \frac{1}{(2 Y )^4}}  & {\scriptstyle \left[  8 Y ^2 \left(4 Y ^2 \left(\mu ^8+8 \mu ^6+30 \mu ^4+8 \mu ^2+1\right)-8 Y  \left(\mu ^5+\mu ^3\right)-\mu ^4\right) I_0(4 Y )^2 \right. }\\ & {\scriptstyle \left.  -8 Y  \mu  \left(4 Y  \left(\mu ^2+1\right)+\mu \right) I_0(4 Y ) \left(\left(4 Y ^2 \left(\mu ^4+6 \mu ^2+1\right)-\mu ^2\right) I_1(4 Y )  +2 Y  \mu \left(4 Y  \left(\mu ^2+1\right)+\mu \right) I_2(4 Y )\right) \right. } \\ & {\scriptstyle \left. -2 \left(16 Y ^4 \left(\mu ^2-1\right)^4-128 Y ^3 \left(\mu ^5+\mu ^3\right)+8 Y^2 \left(\mu ^6+4 \mu ^4+\mu ^2\right)+\mu ^4\right) I_1(4 Y )^2   \right. } \\ & {\scriptstyle \left.  -8 Y ^2 \mu ^2 \left(4 Y  \left(\mu ^2+1\right)+\mu \right)^2 I_2(4 Y )^2  \right. } \\ & {\scriptstyle \left.  +8 Y  \mu  \left(16 Y ^3 \left(\mu ^6+7 \mu ^4+7 \mu ^2+1\right)+4 Y ^2 \left(\mu ^5+6 \mu ^3+\mu \right)+4 Y  \left(\mu ^4+\mu ^2\right)+\mu ^3\right)  I_1(4 Y ) I_2(4 Y )  \right] } \end{aligned} $  \\
				 \hline
				 2 & 3 &   $  \begin{aligned} {\scriptstyle \frac{1}{(2 Y )^6} } & {\scriptstyle \left[   8 Y ^2 \left(-4 Y ^2 \left(27 \mu ^4+62 \mu ^2+27\right) \mu ^4-16 Y ^3  \left(5 \mu ^6-21 \mu ^4-21 \mu ^2+5\right) \mu ^3 \right.  \right. } \\ & {\scriptstyle \left.  +16 Y ^4 \left(\mu ^{12}+21 \mu ^{10}+195 \mu ^8+334 \mu ^6+195 \mu ^4+21 \mu ^2+1\right)-36 Y  \left(\mu ^7+\mu ^5\right)-9 \mu ^6\right) I_0(4 Y )^2 }  \\ &  {\scriptstyle -8 Y  \mu  I_0(4 Y ) \left(2 Y  \mu  \left(4 Y ^2 \left(3 \mu ^4+10 \mu ^2+3\right)+6 Y  \left(\mu ^3+\mu \right)+3 \mu ^2\right)^2 I_2(4 Y )  \right. }\\ &  {\scriptstyle  +\left(32 Y ^5 \left(3 \mu ^{10}+55 \mu ^8+198 \mu ^6+198 \mu ^4+55 \mu ^2+3\right)+16 Y ^4 \mu  \left(3 \mu ^8+60 \mu ^6+130 \mu ^4+60 \mu ^2+3\right) \right. } \\ & {\scriptstyle \left. \left. -32 Y ^3 \mu ^2 \left(2 \mu ^6+3 \mu ^4+3 \mu ^2+2\right)-8 Y ^2 \mu ^3 \left(9 \mu ^4+22 \mu^2+9\right)-36 Y  \left(\mu ^6+\mu ^4\right)-9 \mu ^5\right) I_1(4 Y )\right) } \\ & {\scriptstyle -2 \left(4 Y ^2 \left(9 \mu ^4+26 \mu ^2+9\right) \mu ^4+96 Y ^4 \left(\mu ^2-1\right)^2 \left(\mu ^4+4 \mu ^2+1\right) \mu ^2+64 Y ^6 \left(\mu^2-1\right)^6-2048 Y ^5 \left(\mu ^3+\mu \right)^3  \right. } \\ & {\scriptstyle \left. +48 Y ^3 \left(\mu ^9+11 \mu ^7+11 \mu ^5+\mu ^3\right)+36 Y  \left(\mu ^7+\mu ^5\right)+9 \mu ^6\right) I_1(4 Y )^2 } \\ & {\scriptstyle -8 Y ^2 \mu ^2 \left(4 Y ^2 \left(3 \mu ^4+10 \mu ^2+3\right)+6 Y  \left(\mu ^3+\mu \right)+3 \mu ^2\right)^2 I_2(4 Y )^2 } \\ & {\scriptstyle +8 Y  \mu  \left(32 Y ^5 \left(3 \mu ^{10}+55 \mu ^8+198 \mu ^6+198 \mu ^4+55 \mu ^2+3\right)+16 Y ^4 \mu  \left(3 \mu ^8+60 \mu ^6+130 \mu ^4+60 \mu ^2+3\right)  \right. } \\ &  {\scriptstyle \left.  +96 Y ^3 \left(\mu ^8+8 \mu ^6+8 \mu ^4+\mu ^2\right)+24 Y ^2 \mu ^3 \left(3 \mu ^4+10 \mu ^2+3\right)+36 Y  \left(\mu ^6+\mu ^4\right)+9\mu ^5\right) I_1(4 Y ) \left.  I_2(4 Y )   \right]  } \end{aligned} $  \\
				 \hline
				 3 & 1 &   $ \begin{aligned}  & {\scriptstyle  6 \left(-\left(\mu ^2 (-I_2(6 Y ))+\left(\mu ^2+1\right) I_0(6 Y )-\mu  I_1(6 Y )+\mu  I_3(6 Y )-I_2(6 Y )\right)  \left(\left(\mu ^2+1\right) I_1(6 Y )-\mu  (I_0(6 Y )+I_2(6 Y ))\right)^2   \right. } \\ & {\scriptstyle  \left.  +\left(\left(\mu ^2+1\right) I_0(6 Y )-2 \mu  I_1(6 Y )\right) \left(\left(\left(\mu ^2+1\right) I_0(6 Y )-2 \mu  I_1(6 Y )\right)^2-\left(\left(\mu ^2+1\right) I_1(6 Y )-\mu  (I_0(6 Y )+I_2(6 Y ))\right)^2\right) \right. } \\ & {\scriptstyle  \left.  +\left(\left(\mu ^2+1\right) I_2(6 Y )-\mu  (I_1(6 Y )+I_3(6 Y ))\right)  \left(\left(\left(\mu ^2+1\right) I_1(6 Y )-\mu  (I_0(6 Y )+I_2(6 Y ))\right)^2   \right. \right. } \\ & {\scriptstyle  \left.\left.  -\left(\left(\mu ^2+1\right) I_0(6 Y )-2 \mu  I_1(6 Y )\right) \left(\left(\mu ^2+1\right) I_2(6 Y )-\mu  (I_1(6 Y )+I_3(6 Y ))\right)\right)\right) } \end{aligned}  $  \\
				 \hline
				\end{tabular}
				\caption{Analytic evaluation of $\mz_{U(N)} ^{+1,K}$ in terms of modified Bessel functions for various $N$ and $K$.}
				\label{tab:ZexactNK}
				\end{table}\par

			\subsection{Large $N$ limit: Gapped solutions}
			\label{app:derivomega}

			 		In this appendix we sketch the computation of $\omega (z)$, defined in \eqref{def:resolvent}, which allows to extract the eigenvalue density in the phases with one or more gaps. The procedure is standard and we follow closely \cite{Jain:2013py,Santilli:2020trf}, glossing over many details. We work in Phase Ia, since all other phases are analyzed in similar fashion.\par
			 		Introduce the function $\varrho (z) $ of complex variable $z\in \C$ such that $\varrho (e^{\ii \theta})= \rho (\theta) $ for $e^{\ii \theta } \in \Gamma$. The saddle point equation \eqref{SPE1} is rewritten as 
			 		\begin{equation}
			 		\label{SPE2}
			 			\mathrm{P} \int_{\Gamma} \frac{\dd w }{2 \pi w} \varrho (w) \frac{ z+w}{z-w}  = W^{\prime} (z) 
			 		\end{equation}
			 		where 
			 		\begin{equation*}
			 			W^{\prime} (z) = - \ii \left[  Y \left( z - \frac{1}{z} \right) - \tau \left(  1+ \frac{ \mu }{z- \mu} + \frac{ \mu^{-1}}{ z- \mu^{-1} } \right) \right] .
			 		\end{equation*}
			 		Equation \eqref{SPE2} is valid for $z \in \Gamma$, and is complemented by the normalization condition 
			 		\begin{equation}
			 		\label{normCplx}
			 			\int_{\Gamma} \frac{\dd w}{ 2 \pi \ii w} \varrho (w) = 1.
			 		\end{equation}
			 		Recall that we have started with a $\mathbb{Z}_2$-symmetric system, invariant under $z \mapsto z^{-1}$ for $z \in \ct$. We will thus find an eigenvalue density with symmetric support, and in particular $\partial \Gamma = \left\{ e^{- \ii \theta_0} ,  e^{ \ii \theta_0 } \right\}$ in a one-cut phase. Then, depending on whether the gap opens at $\theta=\pi$ or $\theta =0$, $\Gamma$ will be the arc on the unit circle connecting $- \theta_0$ to $\theta_0$ or $\theta_0$ to $- \theta_0$, respectively, with orientation always taken counter-clockwise.\par
			 		Recall from the definition \eqref{def:resolvent} that 
			 		\begin{equation*}
			 			\omega_{+} (e^{\ii \theta}) - \omega_{-} (e^{\ii \theta}) = 2 \varrho (e^{\ii \theta}) , \qquad e^{\ii \theta} \in \Gamma .
			 		\end{equation*}
			 		In turn, from the definition of Cauchy principal value and \eqref{SPE2} we immediately get 
			 		\begin{equation}
			 		\label{SPEdisc}
			 			\omega_{+} (e^{\ii \theta}) + \omega_{-} (e^{\ii \theta}) = - 2 \ii W^{\prime} (e^{\ii \theta}) .
			 		\end{equation}
			 		The normalization \eqref{normCplx} and the definition \eqref{def:resolvent} imply that $\omega (z) \to 1$ as $\vert z \vert \to \infty$. We have then reduced the problem of finding the eigenvalue density to the problem of determining the discontinuity of $\omega (z)$ along $\Gamma$, from the knowledge of its regular part and the boundary condition $\omega (z \to \infty) =1$. It is standard procedure to reduce the problem \eqref{SPEdisc} to a discontinuity equation for a new, auxiliary function $\Omega (z)$ related to $\omega (z)$ via 
			 		\begin{equation}
			 		\label{eq:defOmegaz}
			 			\omega (z) = \sqrt{\left(e^{\ii \theta_0} - z \right)  \left(e^{- \ii \theta_0} - z \right)} \Omega (z) .
			 		\end{equation}
			 		We take the square root with positive value, but any potential ambiguity in the intermediate steps and definitions from now on, would drop out from the final answer.\par
			 		Writing 
			 		\begin{equation*}
			 			\left[ \sqrt{\left(e^{\ii \theta_0} - z \right)  \left(e^{- \ii \theta_0} - z \right)} \right]_{\pm} = \left[ \sqrt{z} \cdot \sqrt{2 \cos \theta - 2 \cos \theta_0 } \right]_{\pm} = \mp e^{\ii \theta /2} \sqrt{2 \cos \theta - 2 \cos \theta_0 } 
			 		\end{equation*}
			 		for $z= e^{\ii \theta} \in \Gamma $, we obtain from \eqref{SPEdisc} the discontinuity equation for $\Omega (z)$:
			 		\begin{equation}
			 		\label{SPEdisc2}
			 			\Omega_{+} (e^{\ii \theta}) - \Omega_{-} (e^{\ii \theta}) = 2 e^{- \ii \theta /2 }\frac{ \ii W^{\prime} (e^{\ii \theta}) }{  \sqrt{2 \cos \theta - 2 \cos \theta_0 } } .
			 		\end{equation}\par
			 		For a multi-cut phase, with 
			 		\begin{equation*}
			 			\Gamma \cong \left\{ \theta_{0,-} \le \theta \le \theta_{0,+}   \right\} \cup \left\{ \theta_{1,-} \le \theta \le \theta_{1,+}   \right\}  \cup \dots \cup  \left\{ \theta_{k,-} \le \theta \le \theta_{k,+}   \right\}
			 		\end{equation*}
			 		the procedure is the same, but with $\Omega (z)$ defined via 
			 		\begin{equation*}
			 				\omega (z) = \sqrt{ \prod_{j=0} ^{k} \left(e^{\ii \theta_{j,+} } - z \right)  \left(e^{ \ii \theta_{j,-}} - z \right)} \Omega (z) .
			 		\end{equation*}\par
\begin{figure}[htb]
			 		\centering
			 			\includegraphics[width=0.4\textwidth]{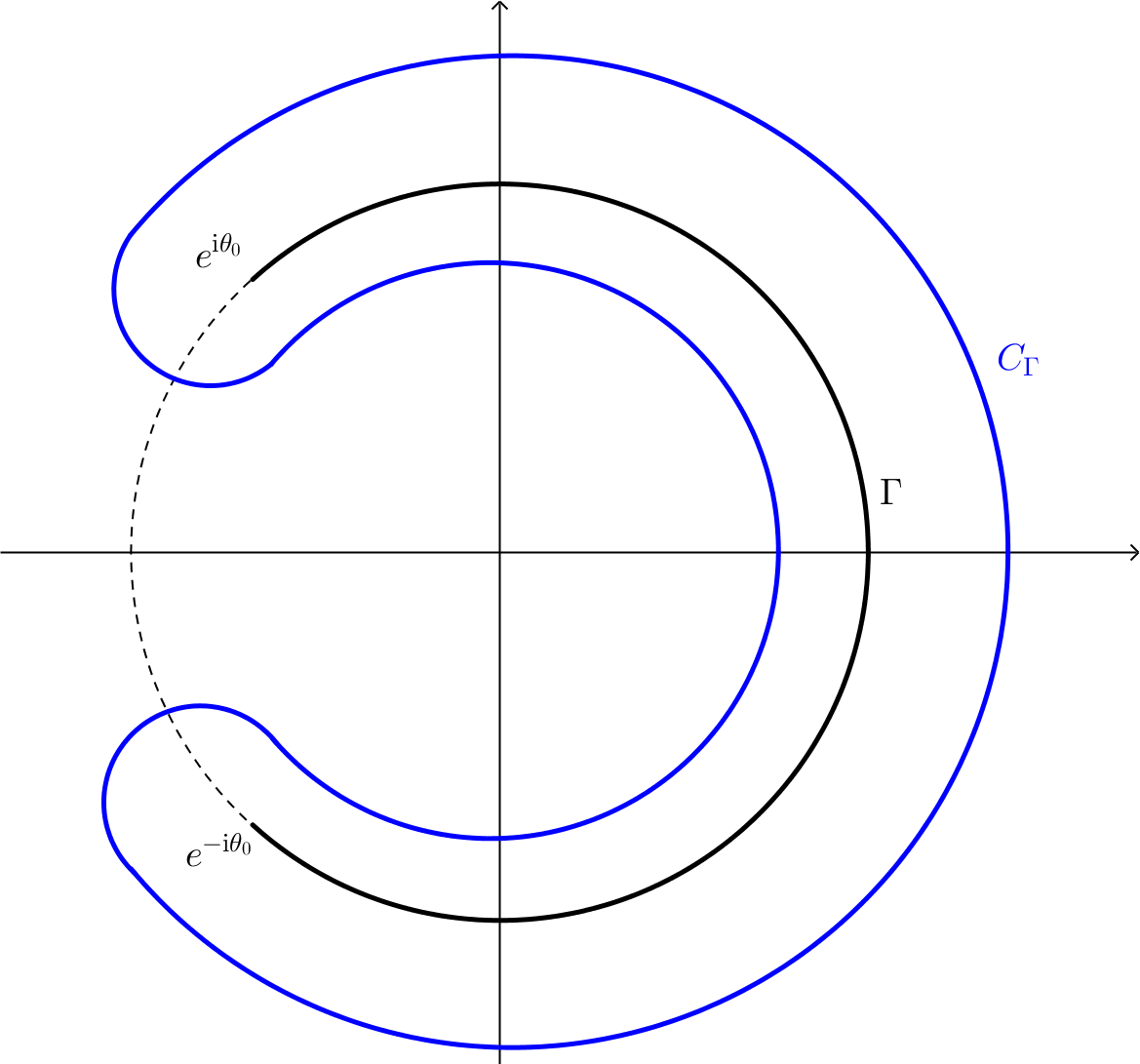}
			 		\caption{Contour $C_{\Gamma}$ encircling the cut $\Gamma$.}
			 		\label{fig:CGamma}
			 		\end{figure}

			 		Let us now introduce a closed contour $C_{\Gamma}$ which is a Jordan curve enclosing $\Gamma$ but not $z$, and oriented counter-clockwise. See Figure \ref{fig:CGamma} for the contour $C_{\Gamma}$ in Phase Ia. 
			 		
			 		\par
			 		
			 		From the definitions \eqref{def:resolvent} and \eqref{eq:defOmegaz} it follows that $\Omega (z)$ falls off (at least) as $1/z$ at infinity. Then, for $z$ lying in the exterior of $C_{\Gamma}$, Cauchy's theorem together with \eqref{SPEdisc2} implies 
			 		\begin{equation*}
			 			\Omega (z)  = \oint_{C_{\Gamma}} \frac{ \dd  w }{ 2 \pi \ii } \frac{ \ii W^{\prime} (w) }{ (z - w) \sqrt{\left(e^{\ii \theta_0} - w \right)  \left(e^{- \ii \theta_0} - w \right) } } 
			 		\end{equation*}
			 		On the other hand, because $W^{\prime} (w)$ is meromorphic we can deform the contour $C_{\Gamma}$ into an infinitely large circle, picking the poles of the integrand. We find 
			 		\begin{align}
			 			\Omega (z) & = - \frac{ \ii W^{\prime} (z)}{  \sqrt{\left(e^{\ii \theta_0} - z \right)  \left(e^{- \ii \theta_0} - z \right) } } - \oint_{C_{\infty}} \frac{ \dd w }{2 \pi \ii } \frac{ \ii W^{\prime} (w) }{ (z - w) \sqrt{\left(e^{\ii \theta_0} - w \right)  \left(e^{- \ii \theta_0} - w \right) } } \notag  \\
			 			& + \sum_{z_{p} \in \left\{ 0, \mu, \mu^{-1}\right\} } \underset{w=z_{p}}{\Res}  \frac{ \ii W (w) }{ (z - w) \sqrt{\left(e^{\ii \theta_0} - w \right)  \left(e^{- \ii \theta_0} - w \right) } } \label{OmegaResolved}
			 		\end{align}
			 		where the first term is the residue at $w=z$, the second term is the remaining contour integral along a circle at infinity, which in our case simply contributes $Y $, and the last term includes the residues at the poles $z_{p}$ of $W^{\prime} (w)$.\par
			 		In Phase Ia, explicit computation of each term leads to 
			 		\begin{equation*}
			 			\omega_{\mathrm{Ia}} (z) = - \ii W^{\prime} (z) + \sqrt{\left(e^{\ii \theta_0} - z \right)  \left(e^{- \ii \theta_0} - z \right) } \left[ Y  \left( 1+ \frac{1}{z} \right) - \frac{\tau }{ \sqrt{1 + \mu^2 - 2 \mu \cos \theta_0} } \left( \frac{\mu }{z- \mu}  -  \frac{1}{z- \mu^{-1}}  \right) \right] .
			 		\end{equation*}
			 		The solution in the other phases is found likewise.

			 	\section{Filling fraction fluctuations}
			 	\label{app:fillingfrac}
			 		
			 		This appendix contains the analysis of the effect of taking into account fluctuations of the filling fractions around the equilibrium configuration.\par
			 		For a generic matrix model in a two-cut phase, the dependence of the filling fractions on the parameters of the theory should be taken into account when computing physical observables \cite{Bonnet:2000dz}.\footnote{The original work \cite{Bonnet:2000dz} dealt with Hermitian matrix models, but the argument extends to the present setting.} Below we briefly review how this effect comes about, and study it for the model at hand. We start with Phase III, and look at Phase II projected onto the real line, as in Section \ref{sec:stereo}.\par

			 			\subsection{Phase III}
			 			\label{app:fffPhIII}
			 				For the two-cut solution in Phase III, let $N_L$ be the number of eigenvalues in the left arc around $\theta=\pi$, $0 \le N_L \le N$, and $N_R = N -  N_L$ the number of eigenvalues in the right arc around $\theta=0$. Let also $\xi= \frac{N_L}{N}$ and $1-\xi= \frac{N_R}{N}$ denote the corresponding filling fractions.\par
			 				The values of $y_L = \cos \left( \theta_{\ast} + \delta \theta \right)$ and $y_R = \cos \left( \theta_{\ast} - \delta \theta \right)$ can be fixed, as functions of $\xi$ and of the other parameters, through the equations 
			 				\begin{align*}
			 					\frac{2}{\pi} \int_{-1} ^{y_L} \dd y \sqrt{  \frac{(y_R -y)(y_L -y )}{ 1-y^2} } \left[  -Y + \frac{ \tau \mu (\mu^2 -1)}{  \sqrt{ (1 + \mu^2 - 2 \mu y_L) (1 + \mu^2 - 2 \mu y_R) } (1 + \mu^2 - 2 \mu y ) } \right] & = \xi  , \\
			 					\frac{2}{\pi} \int_{y_R} ^{1} \dd y \sqrt{  \frac{(y-y_R)(y -y_L )}{ 1-y^2} } \left[  -Y + \frac{ \tau \mu (\mu^2 -1)}{  \sqrt{ (1 + \mu^2 - 2 \mu y_L) (1 + \mu^2 - 2 \mu y_R) } (1 + \mu^2 - 2 \mu y ) } \right] & = 1 - \xi  , 					
			 				\end{align*}
			 				that come from the definition of $\xi$ after changing variables $y= \cos \theta$. Then, the value of $\xi$ is fixed by the equilibrium condition 
			 				\begin{equation}
			 				\label{eq:xisp}
			 					\left. \frac{\dd S_{\mathrm{eff}}}{ \dd \xi } \right\rvert_{\xi = \xi_{\mathrm{sp}} } = 0.
			 				\end{equation}
			 				For example, approximating close to the critical surface diving Phase III from Phase Ib, we find 
			 				\begin{equation*}
			 					\left. \frac{ \partial \xi_{\mathrm{sp}}}{ \partial y_R } \right\rvert_{y_R=1} = \frac{\sqrt{2 - 2y_L}}{\pi} \left( \frac{1}{2} - \frac{\tau}{\mu-1} + \frac{ \tau \mu (\mu+1)}{ (\mu-1)^3  \sqrt{1 + \mu^2 - 2 \mu y_L} }  \right) 
			 				\end{equation*}
			 				where we have also substituted $Y= Y_{\mathrm{cr,b}}$. It has been shown in \cite{Bonnet:2000dz} that the quantum fluctuations around the saddle point $\xi_{\mathrm{sp}}$ contribute to the free energy a term of the form $- \frac{1}{N^2} \log \vartheta (N \xi_{\mathrm{sp}})$, where $\vartheta (z)$ is the Jacobi theta function. See Appendix \ref{app:fffPhII} below for more details and a very short review of the derivation. This is a sub-leading contribution to the free energy but, due to the dependence on $N \xi_{\mathrm{sp}}$, each derivative generates a factor of $N$. Therefore, the $\xi_{\mathrm{sp}}$-dependent part becomes of the same order as the leading order term when differentiating the Wilson loop vev, and must be taken into account. The relevant part of the derivative is 
			 				\begin{equation*}
			 					\sum_{y \in \left\{ y_L, y_R \right\} }  \left[ \frac{ \dd \ }{ \dd z } \log \vartheta (z) \vert_{z= N \xi_{\mathrm{sp}}}  \right]^{2} \left( \frac{\partial y}{\partial Y} \frac{\partial \xi_{\mathrm{sp}}}{ \partial y}  \right)^2  ,
			 				\end{equation*}
			 				which yields a non-trivial contribution to the derivative of the Wilson loop vev in Phase III. However, when approaching the critical loci, $\xi \to 0$ if $Y \to Y_{\mathrm{cr,a}}$ or $\xi \to 1$ if $Y \to Y_{\mathrm{cr,b}}$, and the derivative of the theta function evaluated at an integer vanishes.\par
			 				This shows that the effect of the filling fractions does not play any role in determining the order of the phase transition, despite being non-trivial in the bulk of Phase III.
			 				
			 			\subsection{Phase II}
			 			\label{app:fffPhII}
			 						We now discuss the same effect in Phase II. It is more convenient and akin to the work \cite{Bonnet:2000dz} to do this in the alternative, Hermitian matrix model presentation of Section \ref{sec:stereo}. The argument can be succinctly summarized as follows.\par
			 						Consider a two-cut solution with support $\text{supp} \rho_{\mathrm{II}} ^{\mathrm{p.}} = \Gamma_{L} \cup \Gamma_R$, and denote by $\xi = \frac{N_L}{N}$ and $1- \xi= \frac{N_R}{N}$ the corresponding filling fractions, as above. The saddle point value $\xi_{\mathrm{sp}}$ of $\xi$ is fixed by \eqref{eq:xisp}. Then, in the large $N$ approximation, the partition function takes the form \cite{Bonnet:2000dz} 
			 						\begin{equation*}
			 							\mz = \sum_{N_L =0} ^{N}  e^{- N^2 \mf_{\mathrm{pert}}  - \frac{ N^2}{2} (\xi - \xi_{\mathrm{sp}} )^2 \left. \partial^2 _{\xi} S_{\mathrm{eff}} \right\rvert_{\xi = \xi_{\mathrm{sp}}} + \mathcal{O} ((\xi - \xi_{\mathrm{sp}})^3 ) } 
			 						\end{equation*}
			 						where $\mf_{\mathrm{pert}}$ is the perturbative free energy to all orders in the $\frac{1}{N^2}$ expansion. This yields \cite{Bonnet:2000dz,Claeys:2014jfa}
			 						\begin{equation*}
			 							- \frac{1}{N^2} \log \mz = \mf + \frac{1}{N^2} \mf^{\mathrm{nlo}} - \frac{1}{N^2} \log \vartheta (N \xi_{\mathrm{sp}} ) + \frac{1}{2N^2} \log \left( 2 \pi \left. \partial^2 _{\xi} S_{\mathrm{eff}} \right\rvert_{\xi = \xi_{\mathrm{sp}}}  \right)+ \mathcal{O} (N^{-4}) ,
			 						\end{equation*}
			 						where $\mf$ is the leading order or planar free energy, $\mf^{\mathrm{nlo}}$ the next-to-leading order correction, and (after an implicit resummation) we have recognized the Jacobi theta function $\vartheta (N \xi_{\mathrm{sp}} ) $. The modular parameter of the theta function is $\ii 2 \pi / (\partial_{\xi} ^2 S_{\mathrm{eff}} (\xi_{\mathrm{sp}}))$, and the dependence on it is kept implicit in the notation.\par
			 						For the case at hand, however, the effective action is an even function, the two wells have identical depth, and all the physical observables we consider preserve this property. We thus have $\xi_{\mathrm{sp}}= \frac{1}{2}$, independent of the parameters of the theory, and the effect we have just described will remain sub-leading \cite{Claeys:2014jfa}. This would not be the case for other type of physical observables that are not protected by the parity symmetry. See \cite{Bonnet:2000dz} for discussion and examples.

		\end{appendix}

\bibliography{LatticeRMT}

\end{document}